\documentclass[journal]{IEEEtran}
\usepackage{latexsym}
\usepackage{graphicx}
\usepackage{amsfonts,amssymb,amsmath}
\usepackage{algorithm,algorithmic}
\usepackage{varwidth}
\usepackage{hyperref}

\usepackage[T1]{fontenc}
\usepackage{cite}
\usepackage{subcaption}
\usepackage{comment}
\usepackage{amsthm}
\usepackage{overpic}
\usepackage{steinmetz}
\usepackage{array}
\usepackage{url}
\usepackage{xcolor}
\IEEEoverridecommandlockouts

\theoremstyle{plain}

\newtheorem{remark}{Remark}
\newtheorem{proposition}{Proposition}

\newcommand{\argmax}[1]{{\underset{{#1}}{\mathrm{arg\,max}}}}
\newcommand{\argmin}[1]{{\underset{{#1}}{\mathrm{arg\,min}}}}
\newcommand{\vect}[1]{\mathbf{#1}}

\def\diag{\mathrm{diag}}

\def\kron{\otimes}

\def\Htran{\mbox{\tiny $\mathrm{H}$}}
\def\Ttran{\mbox{\tiny $\mathrm{T}$}}
\def\CN{\mathcal{N}_{\mathbb{C}}} 

\def\mod{\mathrm{mod}}

\def\H{\textrm{H}}
\def\T{\textrm{T}}

\begin{document}

\title{Parametric Channel Estimation with Short Pilots in RIS-Assisted Near- and Far-Field Communications}

\author{Mehdi Haghshenas, Parisa Ramezani, 
Maurizio Magarini, 
Emil Bj{\"o}rnson \\
\thanks{This article has been presented in part in IEEE International Conference on Communication (ICC) 2023 \cite{haghshenas2023efficient}.}%
\thanks{M. Haghshenas and M. Magarini are with the Department of Electronics, Information and Bioengineering, Politecnico di Milano, 20133 Milan, Italy. Email: \{mehdi.haghshenas, maurizio.magarini\}@polimi.it}%
\thanks{E.~Bj\"ornson and P. Ramezani are with the Department of Computer Science, KTH Royal Institute of Technology, SE-100 44 Stockholm, Sweden. Email: \{parram, emilbjo\}@kth.se}%
\thanks{E.~Bj\"ornson and P. Ramezani are supported by the FFL18-0277 grant from Swedish Foundation for Strategic Research. The work of M. Magarini is supported by the European Union under the Italian National Recovery and Resilience Plan (PNRR) of NextGeneration EU, partnership on ``Telecommunications of the Future'' (PE00000001 - program ``RESTART'', Structural Project SRE).
 }
}

\maketitle

\begin{abstract}
Considering the dimensionality of a typical reconfigurable intelligent surface (RIS), channel state information acquisition in RIS-assisted systems requires lengthy pilot transmissions. 
Moreover, the large aperture of the RIS may cause transmitters/receivers to fall in its near-field region, where both distance and angles affect the channel structure. This paper proposes a parametric maximum likelihood estimation (MLE) framework for jointly estimating the direct channel between the user and the base station (BS) and the line-of-sight channel between the user and the RIS, in both far-field and near-field scenarios. The MLE framework is first developed for the case of single-antenna BS and later extended to the scenario where the BS is equipped with multiple antennas. A novel adaptive RIS configuration strategy is proposed to select the RIS configuration for the next pilot to actively refine the estimate. We design a minimal-sized codebook of orthogonal RIS configurations to choose from during pilot transmission with a dimension much smaller than the number of RIS elements. To further reduce the required number of pilots, we propose an initialization strategy with two wide beams. We demonstrate numerically that the proposed MLE method needs only a few pilots for achieving accurate channel estimates and further show that the presented framework performs well under Rician fading.  
We also showcase efficient user channel tracking in near-field and far-field scenarios. 
\end{abstract}

\maketitle

\section{Introduction}
The ever-increasing number of mobile devices and the emergence of new data-intensive applications create the need for new technologies that can cope with the exponentially growing demand for data traffic at a reasonable complexity and cost. 
Recently, the reconfigurable intelligent surface (RIS) technology has appeared with unprecedented features that can fundamentally change how future wireless communication networks are deployed and operated. RIS is an electromagnetic (EM) metasurface consisting of a large number of low-cost passive elements that can dynamically change the properties (e.g., phase) of the impinging waves and thereby reflect them in desired directions \cite{Huang2018,RuiTutorial}. In this way, RIS improves the link quality between communicating devices, which is of particular importance when the direct link between the transmitter-receiver pair is  degraded \cite{Emil2022}.

Although the benefits of RIS-assisted communications have been demonstrated both theoretically and experimentally \cite{9206044,Pei2021a}, the RIS technology faces some major practical challenges that must be overcome before the technology can be widely adopted. Specifically, the professed real-time reconfigurablity of a RIS can only be achieved if the channel state information (CSI) is updated at the same pace as the channels are changing. Yet, CSI acquisition in RIS-assisted communication systems is far from trivial and is claimed to be a critical research challenge to address \cite{9311936}. Due to the lack of active radio frequency components and computational resources, the RIS cannot transmit/receive pilot symbols and perform channel estimation calculations. Hence, the transmitter must send repeated pilots while the RIS sweeps through a predefined set of configurations so that the receiver can observe all possible channel dimensions.
Unfortunately, the huge dimensionality of the RIS makes the matter more difficult as the number of coefficients to be estimated is very large, requiring lengthy training periods \cite{Swindlehurst2022}. 

The real-time reconfigurability of the RIS is even more challenging under user mobility as the CSI acquisition must be performed with higher frequency. The dynamic RIS reconfiguration under user mobility has been considered in several recent works, focusing on scenarios with stationary fading processes. The Jakes’ model was used in \cite{Papazafeiropoulos2022a,Zhang2022a} to describe the temporal fading correlation over non-line-of-sight (NLOS) channels with Rayleigh fading and known statistics.
Line-of-sight (LOS) channels were treated in \cite{Xu2022,Chen2022a} using Rician fading models with similar temporal correlation and LOS paths that are constant over time except for the common phase shifts.
These approaches can track the channel aging for users moving in a tiny region where the channel statistics can be modeled as stationary but cannot manage large-scale mobility (e.g., vehicles or people passing by a RIS). Nevertheless, the use of the approaches in \cite{Papazafeiropoulos2022a,Zhang2022a,Xu2022,Chen2022a} is problematic since a pilot length equal to the number of RIS elements is required. We thus need efficient solutions to reduce the channel estimation overhead in RIS-assisted systems.

One way to reduce the training overhead is to group adjacent RIS elements into a subarray where all the elements share the same reflection coefficient \cite{BZhang2020,RuiHierar2020}. Although this method can reduce the training overhead, it reduces the beamforming gain proportionally to the size of the subarray, and the overhead still scales with the number of subarrays. 
It was shown in \cite{Ozlemarraygeometry} that the spatial channel correlation induced by the array geometry can be exploited to make the pilot length proportional to the RIS area instead of the number of elements. However, this property only makes the pilot overhead over a RIS with many tiny elements identical to that of a RIS with fewer half-wavelength spaced elements, so the pilot overhead remains high for physically large RISs.

A RIS is mainly useful when the direct link between the transmitter and receiver is weak \cite{Emil2022}, a scenario that currently limits the coverage range in high frequency bands (e.g., millimeter wave bands) because the propagation conditions are harsh in NLOS scenarios. A potential solution is to deploy a RIS for range extension, particularly at locations where it has LOS to the potential user locations and good channel conditions to the BS.
Exploiting the fact that channels in these bands can be represented by a few resolvable paths, the structure of the channel can be leveraged to reduce the training overhead. Specifically, if the channels are determined by a small number of parameters (e.g., gains and angles), the required pilot length can ideally be made proportional to that number.  For instance, the authors in \cite{Sepideh2021} exploit the low-rank property of the mmWave channel to convert the cascaded channel estimation problem into a set of angle of arrival (AoA) and angle of departure (AoD) estimations. In \cite{Hongbin2020}, Wang \textit{et al.} utilize the Khatri-Rao and Kronecker products and convert the channel estimation into a sparse recovery problem. The rank-deficient structure of the channel is utilized in\cite{Yuan2019} to transform channel estimation into a joint sparse matrix factorization and matrix completion problem. In \cite{Henk2021}, channel estimation is broken into two compressive sensing (CS) problems and atomic norm minimization is employed to find the angular parameters. A parametric maximum likelihood estimator (MLE) is proposed in \cite{Bjornson2022MLE} for estimating a far-field free-space LOS channel between the user and the RIS when the direct link is blocked.

A large number of RIS elements is needed in practice to compensate for the double attenuation faced by the signal travelling through the RIS \cite{Emil2020RISvsRelay}. The evolution towards higher frequencies and larger arrays will cause future communications to operate in the radiative near-field regime, where previously negligible physical phenomena dominate. Particularly, the planar approximation of the EM wavefronts is no longer valid in the near-field of the RIS, and the EM propagation must be characterized by the more accurate spherical wave models \cite{ramezani2022bookchapter,Direnzo2023}. Consequently, the distance between the RIS and the transmitter/receiver becomes an essential parameter in the modeling near-field channels that must be estimated along with angular parameters when considering channels with few paths. 

There are some recent attempts to develop near-field channel estimation methods for extremely large-scale multiple-input multiple-output (XL-MIMO) systems. In \cite{HanXLestimateSub}, a subarray-based channel estimation technique is presented for the case where an XL-MIMO transmitter communicates with a single-antenna user. The entire array at the transmitter is divided into multiple subarrays, where one or multiple subarrays are visible to the user. This introduces sparsity in the channel structure and a refined orthogonal matching pursuit (OMP) algorithm is utilized for estimating the channel of each subarray. The authors in \cite{LingLongXLestimate} represent the channel in the polar domain where the sparsity of a near-field channel is visible. Then, CS techniques are applied to estimate angular and distance parameters. Reference \cite{Lu2023Mixed} considered a near-field channel model where the LOS and NLOS components are separately modeled. The channel estimation problem is then decomposed into two subproblems, where the estimation of the LOS part is performed by geometric parameter estimation, while the NLOS term is estimated using an OMP-based method. In the context of RIS-assisted communication, Wu \textit{et al.} exploit the polar near-field sparsity to transform the channel estimation into a sparse recovery problem which is then solved by an orthogonal least square algorithm \cite{ShimNear2022}.

One of the main issues with the works on channel estimation for RIS-assisted systems is that the estimation of the direct channel between the transmitter and the receiver is often neglected. It is common to assume that the direct channel can be estimated via conventional estimation techniques by turning the RIS off. However, this strategy does not account for the estimation errors of the direct channel in estimating the cascaded channel, which degrades estimation quality of the latter. To accurately estimate both the direct and cascaded channels, it is important to perform a joint estimation of them. Besides, near-field channel estimation for RIS-assisted systems is still in its very early stage and more investigations are needed to develop efficient channel estimation methods that work effectively both in near- and far-field scenarios. 

\subsection{Contributions}
This paper develops an efficient parametric MLE framework for estimating the channels from the user in a RIS-assisted communication system. Noticeably, the framework can be used in both near-field and far-field scenarios.
To avoid sending pilots using RIS configurations that are unlikely to match the channel, we propose an adaptive approach where the next configuration to be used during pilot transmission is selected based on pilot signals received so far. 
The pilots are selected from a novel codebook with a dimension much lower than the number of RIS elements, which we design utilizing the fact that the RIS channels reside in a low-dimensional subspace. We also propose an efficient way to initiate the MLE algorithm using two wide beams. 
The contributions of the paper can be summarized as follows:

\begin{itemize}
    \item We propose a parametric MLE algorithm for estimating the LOS user-RIS channel as well as the direct channel which recovers the distance from the user to the RIS, along with the gain and angular parameters of the  user-RIS and direct channels, considering a single-antenna BS setup. The proposed framework can be applied to estimate both near-field and far-field channels. We further devise an adaptive RIS configuration strategy which refines the estimation iteratively, thus enabling an accurate estimation of the channels within a handful number of pilot symbols.
    \item We design an efficient codebook for RIS configurations to be used during the channel estimation by utilizing the fact that the channels involving RIS approximately lie in a low-dimensional subspace. The designed codebook has much lower dimension than the prevalent discrete Fourier transform (DFT)-based codebook as it only explores the important dimensions.
    \item We propose an efficient initialization for the MLE algorithm that sends two wide beams covering the left and right sides of the RIS, respectively. This ensures that useful information is collected irrespective of the user location.
    \item We extend our MLE framework to the multi-antenna BS scenario and derive the parameter estimator in this more complicated setup. We further enhance the adaptive RIS configuration strategy for the case of multi-antenna BS.
    \item We demonstrate the effectiveness of our proposed designs and algorithms numerically.  We first show that the MLE algorithm requires few pilots, particularly with the proposed initial RIS configuration, in both far-field and near-field scenarios. We also show that the MLE algorithm is capable of tracking channel variations in the presence of non-stationary user mobility. We further validate the effectiveness of the proposed algorithm under Rician fading channel models. Finally, we compare the adaptive RIS configuration strategy with the widely-used hierarchical beam training method and show that the proposed strategy is notably superior to the hierarchical training although both methods are built on the same principle of adaptively locating the user.     
\end{itemize}

This paper extends the conference version \cite{haghshenas2023efficient} by considering scenarios where the user falls in the near-field of the RIS, making the distance between RIS and the user an essential parameter to be estimated. We further develop a novel initialization strategy for the proposed MLE algorithm, which reduces the number of pilots compared to the case where the algorithm is randomly initialized as in \cite{haghshenas2023efficient}. We also extend our analyses to the case of multi-antenna BS and derive the parameter estimations for this more complicated scenario.

\subsection{Organization and Notations}
The remainder of this paper is structured as follows. In Section \ref{sec:sysmod}, we present the system model of the considered RIS-assisted communication system. Sections \ref{sec:MLE} and \ref{sec:adaptiveRIS} describe the proposed MLE algorithm and the adaptive RIS configuration approach. Section \ref{sec:codebook} develops a low-dimensional codebook to be used during  channel estimation and Section \ref{sec:Init} proposes a novel RIS configuration design for initializing the MLE algorithm. Numerical results are presented in Section \ref{sec:numericalRes} and Section \ref{sec:conclusion} concludes the paper. 

\textit{Notations:} Vectors and matrices are
respectively denoted by boldface lowercase and boldface uppercase
letters. $(\cdot)^{\Htran}$, $(\cdot)^{\Ttran}$, and $(\cdot)^*$ denote Hermitian transpose, transpose, and conjugate, respectively. $\diag(a_1,\ldots,a_N)$ is a diagonal matrix having $a_1,\ldots,a_N$ as its diagonal elements. $\mathrm{Re}(\cdot)$, $|\cdot|$, and $\mathrm{arg}(\cdot)$ indicate the real part, modulus, and angle of a complex number. The symbol $\mathbb{C}$ denotes the complex set and the distribution of a circularly symmetric complex Gaussian random
variable with variance $\sigma^2$ is denoted
by $\sim \CN(0, \sigma^2)$, where $\sim$ stands for “distributed as”.   
The Kronecker and Hadamard products are denoted by $\otimes$ and $\odot$, respectively.  $\lfloor \cdot \rfloor$ and $\lceil \cdot \rceil$ are the floor and ceiling operators which round a real argument to its closest smaller and larger integer, respectively. The symbol $\in$ indicates set membership.

\section{System Model}
\label{sec:sysmod}
\begin{figure}[t!]
	\centering
	\begin{overpic}[width=\columnwidth]{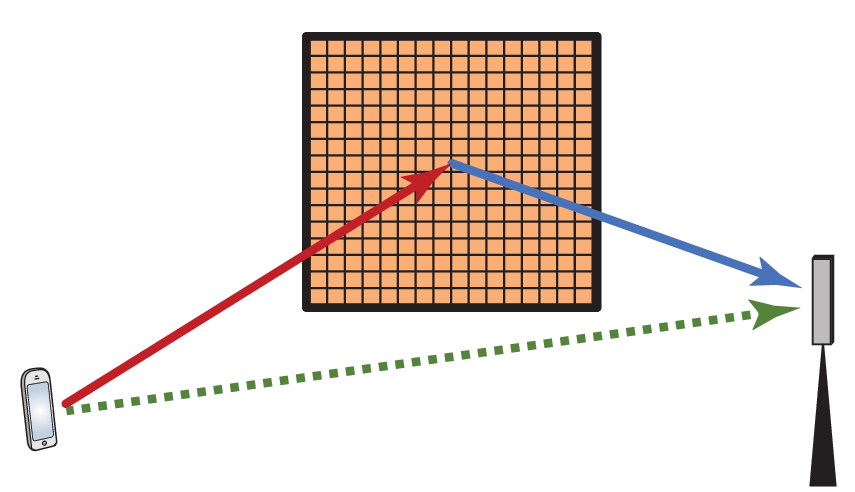}
	 \put (0,1) {Transmitter}
	 \put (79,1) {Receiver}
	 \put (35.3,55) {RIS with $N$ elements}
  \put (9.5,22) {$\vect{g}$ (LOS)}
	 \put (72,33) {$\vect{h}$ (LOS)}
  \put (43,10) {$d$ (NLOS)}
\end{overpic}  
	\caption{A RIS-assisted communication scenario.}
	\label{fig:scenario}
\end{figure}
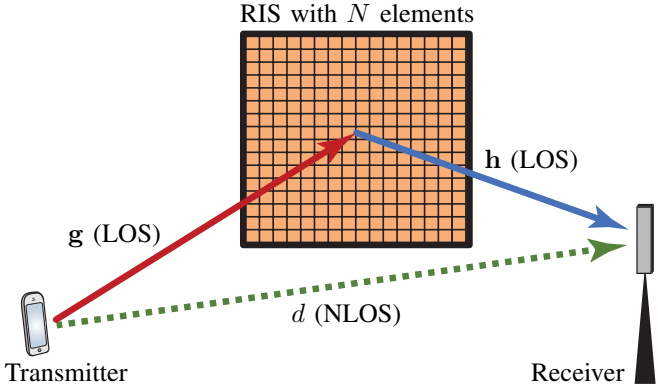

With reference to Figure \ref{fig:scenario}, we consider a communication system where the transmission between a single-antenna base station (BS) and single-antenna users is assisted by a RIS, which consists of $N$ tunable reflecting elements. We denote by $\vect{h}$$\,=\,$$[h_1,\ldots,h_N]^{\Ttran} \in \mathbb{C}^N$ the channel between the BS and the RIS. This channel is  assumed to be static and known because the locations of RIS and BS are fixed. 
The channel between the RIS and an arbitrary user is denoted by $\vect{g} $$\,=\,$$\bigl[g_1,\ldots,g_N\bigr]^{\Ttran} \in \mathbb{C}^N$ and $d \in \mathbb{C}$ represents the direct channel between the BS and the considered user. The values of $\vect{g}$ and $d$ are time-varying due to user mobility; thus, either the BS or the user must estimate the channel periodically and update the RIS phase-shift configuration accordingly to maintain a good SNR. This work focuses on the uplink channel estimation problem, where the user sends a series of pilot symbols, but the same methodology can be used in the downlink. 

Suppose that the user transmits the symbol $x \in \mathbb{C}$. The received signal at the BS can then be expressed as 
\begin{equation}
    y= \left( \boldsymbol{\theta}^{\Ttran} \vect{D}_{\vect{h}} \vect{g} + d \right) x + w,
\end{equation}
where $\boldsymbol{\theta}$$\,=\,$$\left[\theta_1,\ldots,\theta_N\right]^{\Ttran} \in \mathbb{C}^N$ is the RIS phase-shift configuration with $\theta_n = e^{-j\xi_n}$ and $\xi_n$ being the phase shift induced by the $n$th RIS element to the incident signal. Additionally, $\vect{D}_{\vect{h}}$ is the diagonal matrix containing the channel coefficients between the BS and RIS elements, i.e., $\vect{D}_{\vect{h}}= \diag (h_1,\ldots,h_N)$,  and  $w \sim \CN(0,\sigma^2 )$ is the independent complex Gaussian receiver noise. If $x \sim \CN(0,P_{\mathrm{d}})$, the spectral efficiency (SE) for a given RIS configuration can be obtained as \cite{Emil2022}
\begin{align} \label{eq:rate-expression}
    R = &\log_2 \!\left( 1 + \frac{P_{\mathrm{d}} }{\sigma^2}  \Big| \boldsymbol{\theta}^{\Ttran} \vect{D}_{\vect{h}} \vect{g} + d \Big|^2 \right) \notag \\ \leq &\log_2 \!\left( 1 + \frac{P_{\mathrm{d}} }{\sigma^2} \left( \sum_{n=1}^{N} \big|h_n g_n\big| + \big|d\big| \right)^2  \right), 
\end{align}
where the upper bound can be achieved by setting the RIS phase shifts as $\theta_n = \exp \left( -j ( \arg(h_n)+\arg(g_n) - \arg (d) \right)$ for $n=1,\ldots,N$. Since $\vect{h}$ is known, we need to estimate $\vect{g}$ and $d$ to compute the phase-shift configuration that maximizes the SE. To focus on the channel estimation problem, we neglect hardware limitations and assume that $\theta_n$ can attain any value in $[0,2\pi)$.

 We express the direct channel as $d= \sqrt{\alpha}e^{j\vartheta}$  where $\alpha \geq 0$ is the channel gain and  $\vartheta \in [0,2\pi)$ is the phase shift. Further, the RIS-reflected channel $\vect{g}$ is assumed to be LOS-dominant such that 
\begin{align}
\label{eq:A-set}
   \vect{g}\in \mathcal{A} = \big\{ \sqrt{\beta}e^{j\omega}  \vect{a} \bigl({\boldsymbol{\psi}}\bigr): \beta \geq 0,\omega \in [0,2\pi), \boldsymbol{\psi}\in \Psi \big\},
\end{align}
where $\beta$ and $\omega$ respectively represent the channel gain and phase shift to the first element of the RIS. Moreover, $\vect{a}(\boldsymbol{\psi})$ is the array response vector that determines the phase shifts at the remaining elements. It is a function of $\boldsymbol{\psi} = (\varphi, \phi)$ in far-field communication and $\boldsymbol{\psi} = (\varphi, \phi, r)$ if the user is in the radiative near-field region of the RIS. Here, $\varphi$ and $\phi$ are the azimuth and elevation angle-of-arrivals (AoAs), respectively, and $r$ is the distance from the user to the first RIS element. The set $\Psi$ of feasible angles and distances will be specified later. The array response vector for a uniform planar RIS can be expressed as \cite{LingLongXLestimate}
\begin{align}
\label{eq:array_response_general}
  \vect{a}(\boldsymbol{\psi}) = \Big [1, e^{j \frac{2\pi}{\lambda} (r-r_2)}, e^{j \frac{2\pi}{\lambda}(r-r_3)},\ldots,e^{j \frac{2\pi}{\lambda} (r-r_N)}\Big]^{\Ttran},
\end{align}
where $r_n$ denotes the distance between the user and the $n$th element of the RIS and we have $r_1 = r$.
We denote by $N_{\mathrm{H}}$ and $N_{\mathrm{V}}$ the number of elements in each row and column of the RIS, respectively, such that $N = N_{\mathrm{H}}N_{\mathrm{V}}$. We index the elements row-by-row and let $n_{\mathrm{H}}$ and $n_{\mathrm{V}}$ be the horizontal and vertical indices of the $n$th RIS element, respectively, which can be computed as 
\begin{equation}
    n_{\mathrm{H}} = \mod (n-1,N_{\mathrm{H}}) +1,~~n_{\mathrm{V}} = \left\lceil \frac{n}{N_{\mathrm{H}}} \right\rceil.
\end{equation}
Assuming that the reference element of the RIS is placed at Euclidean 3D coordinate $(0,0,0)$ and the user is located at $r(\cos \varphi \cos \phi,\sin \varphi \cos \phi, \sin \phi )$, we have
\begin{align}
\label{eq:r_n2}
    r_n ^2 &= (r\cos \varphi \cos \phi)^2 + (r\sin \varphi \cos \phi - (n_{\mathrm{H}} - 1)\Delta_{\mathrm{H}})^2 \notag \\ &+ (r\sin \phi - (n_{\mathrm{V}} - 1)\Delta_{\mathrm{V}})^2 \notag \\ & = r^2 - 2r \big( (n_{\mathrm{H}} - 1)\Delta_{\mathrm{H}} \sin \varphi \cos \phi + (n_{\mathrm{V}} - 1)\Delta_{\mathrm{V}}\sin \phi  \big) \notag \\
    & + \big( (n_{\mathrm{H}} - 1)\Delta_{\mathrm{H}} \big)^2 + \big( (n_{\mathrm{V}} - 1)\Delta_{\mathrm{V}} \big)^2,
\end{align}where $\Delta_{\mathrm{H}}$ and $\Delta_{\mathrm{V}}$ are the inter-element spacing between adjacent horizontal and vertical RIS elements. Taking the square root of both sides in \eqref{eq:r_n2}, we obtain 
\begin{equation}
    r_n = r \sqrt{1 \!-\! \frac{2\big( i(n) \sin \varphi \cos \phi + k(n) \sin \phi \big)}{r} \!+\! \frac{i(n)^2 + k(n)^2}{r^2}},
\end{equation}where 
\begin{equation}
    i(n) = (n_H -1)\Delta_H,~~k(n) = (n_V -1)\Delta_V.
\end{equation}
Therefore, the distance between the RIS's reference and $n$th element turns out to be given by
\begin{small}
\begin{align}
\label{eq:r_n}
    &r_{1n} = r - r_n =\notag \\
    &r \left( 1 - \sqrt{1 - \frac{2\big( i(n) \sin \varphi \cos \phi + k(n) \sin \phi \big)}{r} + \frac{i(n)^2 + k(n)^2}{r^2}}\right).
\end{align}\end{small}

\subsection*{Special Case: Far-Field}
The array response vector defined above is applicable in the radiative near-field, at distances where the amplitude variations over the array are negligible \cite{ramezani2022bookchapter}. The expression can also be used in the far-field; however, when the user is in the far-field of the RIS, we can simplify the relative distance expression in \eqref{eq:r_n} and obtain an array response vector that is independent of the user's distance to the RIS. Specifically, using the Taylor approximation $\sqrt{1+x} \approx 1 + x/2$, we have 
\begin{equation}
\label{eq:r_n_far}
    r_{1n} \approx i(n) \sin \varphi \cos \phi + k(n) \sin \phi  - \frac{i(n)^2 + k(n)^2}{2r}.
\end{equation}
In the far-field scenario where $r \gg i(n),~r \gg k(n)$, the last term in \eqref{eq:r_n_far} can be neglected and the relative distance can be approximated as $r_{1n} \approx i(n) \sin \varphi \cos \phi + k(n) \sin \phi $. Therefore, the far-field array response vector becomes
\begin{equation}
\label{eq:array_response_farfield}
    \begin{aligned}
        &\mathbf{a}(\varphi,\phi) = \left [ 1,\ldots, e^{j\frac{2\pi}{\lambda}[i(n) \sin (\varphi)\cos (\phi)  + k(n)
        \sin (\phi)]} \right. \\
        & ,\ldots, \left. e^{j\frac{2\pi}{\lambda}[(N_\textrm{H} - 1)\Delta_\textrm{H} \sin (\varphi)\cos(\phi)+ (N_\textrm{V} - 1) \Delta_\textrm{V} \sin (\phi)]} \right ]^{\Htran},
    \end{aligned}
\end{equation}which is independent of the distance $r$.

\section{Maximum Likelihood Channel Estimator}
\label{sec:MLE}
In this section, we develop a parametric maximum likelihood channel estimator for the joint estimation of the channels $\vect{g}$ and $d$. To enable the estimation of these channels, the user transmits a known pilot signal at $L$ time instances and the RIS uses a different configuration for each of them so that the projections of $\vect{g}$ in different directions are observable. Particularly, if the user transmits the deterministic pilot signal $x_{\mathrm{p}} = \sqrt{P_\mathrm{p}}$ with power $P_{\mathrm{p}}$ at $L$ time instances with the RIS being configured as $\boldsymbol{\theta}_1,\ldots,\boldsymbol{\theta}_L$, the concatenated received signal $\vect{y} \in \mathbb{C}^L$ at the BS can be expressed as
\begin{equation} \label{eq:received-pilot}
    \vect{y} = \left(\vect{B} \vect{D}_{\vect{h}} \vect{g} + d \vect{1}_{L \times 1} \right)\sqrt{P_\mathrm{p}} + \vect{w},
\end{equation} where
\begin{align}\label{eq:configuration_matrix} \vect{B} &= \left[\boldsymbol{\theta}_1,\ldots,\boldsymbol{\theta}_L\right]^{\Ttran}, \\
    \vect{w} &= [w_1,\ldots,w_L]^{\Ttran},
\end{align} 
and $w_l \sim \CN(0,\sigma^2 )$ is the independent noise at pilot time instance $l$, for  $l=1,\ldots,L$.

There is a multitude of channel estimators that can be developed based on the received pilot signal in \eqref{eq:received-pilot}.
We will take the MLE approach \cite{Kay1993a} because the channels $\vect{g}$ and $d$ are assumed to be unknown but not having a known random characterization due to non-stationary mobility. 
The probability density function (PDF) of $\vect{y}$ for given values of $\vect{g}$ and $d$ can be expressed as 
\begin{equation}
\label{eq:y_PDF}
    f_{\vect{Y}}(\vect{y}; \vect{g},d) = \frac{1}{(\pi \sigma ^2)^L} e^{- \frac{\|\vect{y} - \left(\vect{B}\vect{D}_{\vect{h}} \vect{g} + d\vect{1} \right) \sqrt{P_\mathrm{p}}\|^2}{\sigma^2}}.
\end{equation} 
We are looking for the channel estimates $\hat{\vect{g}}$ and $\hat{\vect{d}}$ that maximize the PDF in \eqref{eq:y_PDF}. This is equivalent to minimization of the squared norm in the exponent; hence, we can formulate the MLE problem as
\begin{align}
\label{eq:g-d-ML-estimation}
    &\{\hat{\vect{g}},\hat{d}\} =  \argmin{\vect{g} \in \mathcal{A},d\in \mathbb{C}}~ \left\|\vect{y} - \left(\vect{B}\vect{D}_{\vect{h}} \vect{g} + d\vect{1} \right) \sqrt{P_\mathrm{p}} \right\|^2 \notag \\ & = \argmin{\vect{g} \in \mathcal{A},d\in \mathbb{C}} P_\mathrm{p}\left\|\vect{B D_h g} + d\vect{1} \right\|^2 - 2\sqrt{P_\mathrm{p}} \mathrm{Re} \left(\vect{y}^{\Htran} (\vect{B D_h g} + d \vect{1})\right) \notag \\
    & = \argmin{\vect{g} \in \mathcal{A},d\in \mathbb{C}}P_\mathrm{p} \left\| \vect{BD_h g} \right \|^2 - 2\sqrt{P_\mathrm{p}}\mathrm{Re} \left( \vect{y}^{\Htran} \vect{BD_h g}\right) + P_\mathrm{p} L |d|^2  \notag \\ & - 2\sqrt{P_\mathrm{p}}\mathrm{Re}\left ( d^* \vect{1}^{\Ttran} \left( \vect{y} - \sqrt{P_\mathrm{p}} \vect{BD_h g}\right)\right). 
\end{align}
We will now utilize the decomposition of the direct channel $d = \sqrt{\alpha} e^{j\vartheta}$. These channel parameters only appear in the last two terms of \eqref{eq:g-d-ML-estimation}. Thus,
we first find the estimate of the gain and the phase shift of $d$. Discarding the terms independent of $d$ in \eqref{eq:g-d-ML-estimation}, we arrive at 
\begin{align}
\label{eq:d-estimate}
   & \{\hat{\alpha},\hat{\vartheta}\} = \notag \\ &\argmin{\alpha \geq 0,\vartheta \in [0,2\pi) } P_\mathrm{p} L \alpha - 2\sqrt{P_\mathrm{p} \alpha}   \mathrm{Re}\left(e^{-j\vartheta} \vect{1}^{\Ttran} (\vect{y} - \sqrt{P_\mathrm{p}}\vect{BD_h g}) \right)\!,
\end{align}
where the minimum is obtained when the real function is maximized. The phase shift of the direct channel is thus obtained as 
\begin{equation}
\label{eq:phase-d-estimate}
    \hat{\vartheta} = \arg \left( \vect{1}^{\Ttran} (\vect{y} - \sqrt{P_\mathrm{p}} \vect{BD_h g})\right),
\end{equation}
which makes the argument of $\mathrm{Re}(\cdot)$ real and positive. Substituting \eqref{eq:phase-d-estimate} into \eqref{eq:d-estimate}, the resulting problem for finding the direct channel gain is expressed as
\begin{equation}
  \hat{\alpha} = \argmin{\alpha \geq 0}~ P_{\mathrm{p}} L \alpha - 2\sqrt{P_{\mathrm{p}}} \sqrt{\alpha} \big|  \vect{1}^{\Ttran} (\vect{y} - \sqrt{P_{\mathrm{p}}}\vect{BD_h g}) \big|,  
\end{equation}in which the objective function is a quadratic function of $\sqrt{\alpha}$ with its minimum being given by 
\begin{equation}
    \sqrt{\alpha} = \frac{\big|  \vect{1}^{\Ttran} (\vect{y} - \sqrt{P_{\mathrm{p}}}\vect{BD_h g}) \big|}{\sqrt{P_{\mathrm{p}}}L}.
\end{equation}
Utilizing the parametrization of $\vect{g}$ given in \eqref{eq:A-set} and substituting $\hat{d} = \sqrt{\hat{\alpha}} e^{j\hat{\vartheta}}$ into \eqref{eq:g-d-ML-estimation}, the remaining MLE subproblem can be expressed as 
\begin{align}
\label{eq:g-estimate}
    &\{\hat{\beta},\hat{\omega} ,\hat{\boldsymbol{\psi}}\} = \notag \\ &\argmin{\substack{\beta \geq 0,\omega \in [0,2\pi),\\ \boldsymbol{\psi} \in \Psi}} P_{\mathrm{p}}\beta  \left(\left \| \vect{BD_h a}(\boldsymbol{\psi})\right\|^2  - \frac{1}{L}\left| \vect{1}^{\Ttran} \vect{BD_h a}(\boldsymbol{\psi})\right|^2 \right)  \notag \\
    &-2\sqrt{P_{\mathrm{p}} \beta} \mathrm{Re}\left(e^{j\omega}\vect{y}^{\Htran}\left( \vect{I}_L - \frac{1}{L}\vect{1}_{L \times L} \right)\vect{BD_h}\vect{a}(\boldsymbol{\psi})\right).
\end{align}
We first notice that $\omega$ only appears in the last term in \eqref{eq:g-estimate}, which is minimized when the term inside $\mathrm{Re}(\cdot)$ is real and positive. The phase shift of the user-RIS channel is thus obtained as 
\begin{align}
\label{eq:phase-g-estimate}\hat{\omega} &= - \arg \left(\vect{y}^{\Htran}\left( \vect{I}_L - \frac{1}{L}\vect{1}_{L \times L} \right)\vect{BD_h}\vect{a}(\boldsymbol{\psi})\right). 
\end{align}After substituting \eqref{eq:phase-g-estimate} into \eqref{eq:g-estimate}, we will have 
\begin{align}
\label{eq:g-estimate-2}
    \{\hat{\beta},\hat{\boldsymbol{\psi}}\} &=\argmin{\beta \geq 0, \boldsymbol{\psi} \in \Psi} P_{\mathrm{p}}\beta  \left(\left \| \vect{BD_h a}(\boldsymbol{\psi})\right\|^2  - \frac{1}{L}\left| \vect{1}^{\Ttran} \vect{BD_h a}(\boldsymbol{\psi})\right|^2 \right)  \notag \\
    &-2\sqrt{P_{\mathrm{p}} \beta} \left|\vect{y}^{\Htran}\left( \vect{I}_L - \frac{1}{L}\vect{1}_{L \times L} \right)\vect{BD_h}\vect{a}(\boldsymbol{\psi})\right|,
\end{align}
where the objective function is quadratic with respect to $\sqrt{\beta}$, yielding
\begin{align}
\label{eq:gain-g-estimate}
\sqrt{\beta} &= \frac{1}{\sqrt{P_{\mathrm{p}}}} \frac{\left| \vect{y}^{\Htran}\left( \vect{I}_L - \frac{1}{L}\vect{1}_{L \times L} \right)\vect{BD_h}\vect{a}(\boldsymbol{\psi}) \right|}{\left \| \vect{BD_h a}(\boldsymbol{\psi})\right\|^2  - \frac{1}{L}\left| \vect{1}^{\Ttran} \vect{BD_h a}(\boldsymbol{\psi})\right|^2} . 
\end{align}
After putting  \eqref{eq:gain-g-estimate} into \eqref{eq:g-estimate-2}, we finally have 
\begin{align}
\label{eq:psi-estimate}
    \hat{\boldsymbol{\psi}} &= \argmin{\boldsymbol{\psi} \in \Psi} - \frac{\left| \vect{y}^{\Htran}\left( \vect{I}_L - \frac{1}{L}\vect{1}_{L \times L} \right)\vect{BD_h}\vect{a}(\boldsymbol{\psi}) \right|^2}{\left \| \vect{BD_h a}(\boldsymbol{\psi})\right\|^2  - \frac{1}{L}\left| \vect{1}^{\Ttran} \vect{BD_h a}(\boldsymbol{\psi})\right|^2} \notag \\ &= \argmax{\boldsymbol{\psi} \in \Psi}  \frac{\left| \vect{y}^{\Htran}\left( \vect{I}_L - \frac{1}{L}\vect{1}_{L \times L} \right)\vect{BD_h}\vect{a}(\boldsymbol{\psi}) \right|^2}{\left \| \vect{BD_h a}(\boldsymbol{\psi})\right\|^2  - \frac{1}{L}\left| \vect{1}^{\Ttran} \vect{BD_h a}(\boldsymbol{\psi})\right|^2}. 
\end{align}
The function involved in \eqref{eq:psi-estimate} can be interpreted as a spatial power spectrum that we should find the tallest peak of.
MLE subproblems of the kind may have many local maxima \cite{Krim1996a}. We might identify all of them \cite{Wang2021a} but in our case, a simple grid search is sufficient because we will develop an algorithm in Section \ref{sec:adaptiveRIS} that iteratively improves the utility function in \eqref{eq:psi-estimate} by sending new pilots until the global peak value is easily distinguished from the local ones. We thus solve \eqref{eq:psi-estimate} numerically by performing a grid search over the set of feasible $\boldsymbol{\psi}$'s. 

We have now derived the solution to the MLE problem and thereby proved the following proposition.
\begin{proposition}
    \label{th:main-result}
   The parametric MLE of $d$ and $\vect{g}$ is obtained as $\hat{d} = \sqrt{\hat{\alpha}} e^{j\hat{\vartheta}}$ 
 and $\hat{\vect{g}} = \sqrt{\hat{\beta}} e^{j\hat{\omega}}\vect{a}(\hat{\boldsymbol{\psi}})$ where
\vspace{-0.5mm}
 \begin{align}
 \label{eq:AoA_estimate_2D}\hat{\boldsymbol{\psi}} &= \argmax{\boldsymbol{\psi} \in \Psi}  \frac{\left| \vect{y}^{\Htran}\left( \vect{I}_L - \frac{1}{L}\vect{1}_{L \times L} \right)\vect{BD_h}\vect{a}(\boldsymbol{\psi}) \right|^2}{\left \| \vect{BD_h a}(\boldsymbol{\psi})\right\|^2  - \frac{1}{L}\left| \vect{1}^{\Ttran} \vect{BD_h a}(\boldsymbol{\psi})\right|^2},\\
 \label{eq:phase_g_estimate}\hat{\omega} &= - \arg \left(\vect{y}^{\Htran}\left( \vect{I}_L - \frac{1}{L}\vect{1}_{L \times L} \right)\vect{BD_h}\vect{a}(\hat{\boldsymbol{\psi}})\right), \\
   \label{eq:amp_g_estimate} \hat{\beta} &= \frac{1}{P_{\mathrm{p}}} \left(\frac{\left| \vect{y}^{\Htran}\left( \vect{I}_L - \frac{1}{L}\vect{1}_{L \times L} \right)\vect{BD_h}\vect{a}(\hat{\boldsymbol{\psi}}) \right|}{\left \| \vect{BD_h a}(\hat{\boldsymbol{\psi}})\right\|^2  - \frac{1}{L}\left| \vect{1}^{\Ttran} \vect{BD_h a}(\hat{\boldsymbol{\psi}})\right|^2} \right) ^2,\\    
     \label{eq:phase_d_estimate}\hat{\vartheta} &= \arg \left( \vect{1}^{\Ttran} (\vect{y} - \sqrt{P_\mathrm{p}} \vect{BD_h} \hat{\vect{g}})\right),  \\
    \label{eq:amp_d_estimate}\hat{\alpha} &= \frac{\left| \vect{1}^{\Ttran} \left( \vect{y} - \sqrt{P_\mathrm{p}} \vect{BD_h}\hat{\vect{g}} \right)\right|^2}{P_{\mathrm{p}} L^2}.
 \end{align}
\end{proposition}

The parameter estimates are interconnected in the sense that
we must first compute $\hat{\boldsymbol{\psi}}$ using \eqref{eq:AoA_estimate_2D} and use it to compute the MLE of
$\hat{\vect{g}}$ from \eqref{eq:phase_g_estimate} and \eqref{eq:amp_g_estimate}. Finally, $\hat{\vect{g}}$ is used to compute $\hat{d}$ using \eqref{eq:phase_d_estimate} and \eqref{eq:amp_d_estimate}.
This is the opposite order of how the estimates were derived. Since we derived an MLE, it will find the exact estimate as the SNR goes to infinity, if the element spacing is smaller or equal to $\lambda/2$. The latter condition is necessary to ensure that the array response is unique with respect to its parameters.

We can now configure the RIS based on the estimates. Since there are no channel statistics, the RIS should be configured as if the estimates were perfect. Particularly, the phase shift vector
\begin{equation}
\label{eq:RIS-Config}
\bar{\boldsymbol{\theta}} = e^{j (\hat{\vartheta} - \hat{\omega})}\diag \left(e^{-j\mathrm{arg}(h_1)},\ldots,e^{-j\mathrm{arg}(h_N)} \right) \vect{a}^*(\hat{\boldsymbol{\psi}}),  
\end{equation}
maximizes the SE in \eqref{eq:rate-expression} if the parameters are perfectly estimated. We notice that it depends on $\hat{\boldsymbol{\psi}}$ and phase shifts $\hat{\omega}, \hat{\vartheta}$, but not on the channel gains $\hat{\alpha},\hat{\beta}$ since we only want to align all signal paths in phase.

\section{Adaptive RIS Configuration During Estimation}
\label{sec:adaptiveRIS}
The proposed parametric channel estimator can reduce the number of necessary pilots to the number of parameters, but this pilot reduction is only achieved when the SNR is sufficiently high. Since the pilot SNR depends on the RIS configuration, the selection of RIS configurations during pilot transmission plays a crucial role in determining the quality of the described channel estimation. In this section, we introduce an adaptive RIS configuration strategy designed to achieve many high-SNR observations of the channel during the estimation phase by adjusting the RIS beamforming direction toward different positions resembling the true channel.

The choice of RIS configurations to be used during the transmission of the pilots determines the quality of the channel estimation explained above.
In this section, we propose an adaptive RIS configuration strategy that aims at attaining sufficiently accurate channel estimation using only a few pilots. The proposed strategy reconfigures the RIS dynamically during the channel estimation to gradually refine the estimation accuracy.

The core idea is to select the columns of $\mathbf{B}$ sequentially, based on already obtained channel knowledge.
If we use the subscript $l$ to denote the estimated parameters when $l$ pilots are transmitted, the SE-maximizing RIS configuration can be obtained from \eqref{eq:RIS-Config} as
\begin{equation}
\label{eq:RIS-Config_l}
\bar{\boldsymbol{\theta}}_l \hspace{-.05cm}=\hspace{-.05cm} e^{j (\hat{\vartheta}_l - \hat{\omega}_l)}\diag \hspace{-.05cm}\left(\hspace{-.05cm}e^{-j\mathrm{arg}(h_1)},\ldots,e^{-j\mathrm{arg}(h_N)}\hspace{-.05cm} \right)\hspace{-.05cm} \vect{a}^*(\hat{\boldsymbol{\psi}}_l).  
\end{equation}
We select the next RIS configuration for pilot transmission based on the phase-shift vector in \eqref{eq:RIS-Config_l}. Specifically, we first define a set $\mathcal{B}= \{ \bar{\boldsymbol{\psi}}_1,\bar{\boldsymbol{\psi}}_2,\ldots,\bar{\boldsymbol{\psi}}_N \}$ of plausible points in space to be considered during channel estimation. 
Based on $\mathcal{B}$, the resulting codebook of RIS configurations is then obtained as 
\begin{align}
\label{eq:configuration_set}
 \Theta \hspace{-.04cm}= \hspace{-.04cm}\left\{\hspace{-.04cm}\diag \hspace{-.04cm}\left(\hspace{-.04cm}e^{-j\mathrm{arg}(h_1)},\ldots,e^{-j\mathrm{arg}(h_N)} \hspace{-.04cm}\right)\hspace{-.04cm} \vect{a}^*(\boldsymbol{\psi}) \hspace{-.1cm}: \hspace{-.04cm}\boldsymbol{\psi} \hspace{-.04cm}\in \hspace{-.04cm}\mathcal{B} \hspace{-.04cm} \right \}\hspace{-.04cm}.  
\end{align}
The design of this codebook will be considered in Section~\ref{sec:codebook} with the goal of spanning all likely channel dimensions with a minimum number of configurations.

Our proposed MLE method can be explained as follows: We start with two RIS configurations $\boldsymbol{\theta}_1$ and $\boldsymbol{\theta}_2$ and transmit pilots at two time instances utilizing the selected RIS configuration matrix $\vect{B}_2 = [\boldsymbol{\theta}_1,\boldsymbol{\theta}_2]^{\Ttran}$. We then use Theorem \ref{th:main-result} to estimate the angular parameters and distance from \eqref{eq:AoA_estimate_2D} via a 3D search and obtain $\hat{\omega}$, $\hat{\beta}$, and $\hat{\vartheta}$ from \eqref{eq:phase_g_estimate}-\eqref{eq:phase_d_estimate}. The RIS configuration for the $(l+1)$th pilot transmission is then selected as the unused configuration in $\Theta$ that is closest to $\bar{\boldsymbol{\theta}}_l$ in \eqref{eq:RIS-Config_l}, i.e., the RIS configuration suggested by the latest channel estimate. Specifically, 
\begin{equation}
\label{eq:new_config}
    \boldsymbol{\theta}_{l+1} = \argmax{\boldsymbol{\theta} \in \Theta} \,\, |\bar{\boldsymbol{\theta}}_l^{\Htran} \boldsymbol{\theta}|.
\end{equation}
A new pilot is then transmitted with the RIS using the configuration in \eqref{eq:new_config} and $\boldsymbol{\theta}_{l+1}$ is removed from $\Theta$; the RIS configuration matrix and the received signal vector are updated as $\vect{B}_{l+1} = [\vect{B}_l^{\Ttran},\boldsymbol{\theta}_{l+1}]^{\Ttran}$ and $\vect{y}_{l+1} = [\vect{y}_l^{\Ttran},y_{l+1}]^{\Ttran}$. This process is repeated until the intended number of pilot transmissions $L$ is reached. 

This iterative procedure is summarized in Algorithm~\ref{Alg:ML_Estimation} and is crafted to deal with the ambiguity that exists in the objective function in Step 5. When the number of pilots is small, this function has multiple peaks of similar height because of the spatial ambiguity existing when only a few channel dimensions are excited by pilots. By selecting a configuration that is aligned to the peak that was used to compute $\hat{\boldsymbol{\psi}}_l$, the function can change in two different ways for the next iteration. The peak will disappear if it was created by ambiguity, in which case $\hat{\boldsymbol{\psi}}_{l+1}$ will be very different from $\hat{\boldsymbol{\psi}}_l$. If not, we have found an approximately optimal solution, and all other peaks will diminish, resulting in $\hat{\boldsymbol{\psi}}_{l+1} \approx \hat{\boldsymbol{\psi}}_l$ and convergence. In either case, the estimate is improved monotonically in each iteration and we only need to consider a tiny fraction of $\Theta$.

Note that the proposed adaptive configuration strategy is based on a pre-established codebook, where each column, identified by its index, signifies a specific RIS configuration. To set the RIS configuration after each pilot transmission, the BS only needs to send an index pointing to the corresponding column in the codebook. For a codebook with $N$ columns, the index value can be conveyed using $\lceil\log_2(N)\rceil$ bits. Therefore, some resources are required between the RIS and the BS for index transmission. In Section~\ref{sec:numericalRes}, we demonstrate the significant reduction in pilot overhead achieved by this adaptive method, although at the expense of a few feedback bits.

\begin{algorithm}[t!]
\small
\caption{Parametric MLE of $\vect{g}$ and $d$.}
\label{Alg:ML_Estimation}
\begin{algorithmic}[1]
\STATE \textbf{Input}: $\vect{D_h}= \diag(h_1,\ldots,h_N)$, codebook $\Theta$,  $\{\boldsymbol{\theta}_1,\boldsymbol{\theta}_2\}$, and L
\STATE Set $\vect{B}_2 = [\boldsymbol{\theta}_1,\boldsymbol{\theta}_2]^{\Ttran}$ 
\STATE{Send pilot signals using the RIS configurations $\boldsymbol{\theta}_1,\boldsymbol{\theta}_2$ to get the received signal $\vect{y}_2 \in \mathbb{C}^2$}
 \FOR{$l = 2,\ldots,L$}
    \STATE{Compute $\hat{\boldsymbol{\psi}}_l  =\argmax{\boldsymbol{\psi} \in \Psi}~ \frac{\left| \vect{y}^{\Htran}\left( \vect{I}_l - \frac{1}{l}\vect{1}_{l \times l} \right)\vect{B}_l\vect{D_h}\vect{a}(\boldsymbol{\psi}) \right|^2}{\left \| \vect{B}_l\vect{D_h a}(\boldsymbol{\psi})\right\|^2  - \frac{1}{l}\left| \vect{1}^{\Ttran} \vect{B}_l\vect{D_h a}(\boldsymbol{\psi})\right|^2}$}
    \STATE {Obtain $\hat{\omega}_l$, $\hat{\beta}_l$, and $\hat{\vartheta}_l$ from \eqref{eq:phase_g_estimate}-\eqref{eq:phase_d_estimate}}
    \IF{$l < L$}
    \STATE{Compute $\bar{\boldsymbol{\theta}}_l$ as in \eqref{eq:RIS-Config_l}}
    \STATE{From the codebook $\Theta$, find the RIS configuration that is closest to $\bar{\boldsymbol{\theta}}_l$, i.e.,  $\boldsymbol{\theta}_{l+1} = \argmax{\boldsymbol{\theta} \in \Theta}~ | \bar{\boldsymbol{\theta}}_l^{\Htran} \boldsymbol{\theta}|$}
    \STATE{Set $\vect{B}_{l+1} = [\vect{B}_l^{\Ttran}, \boldsymbol{\theta}_{l+1}]^{\Ttran}$, update $\Theta \gets \Theta \setminus \{\boldsymbol{\theta}_{l+1}\}$}
    \STATE{Send a pilot signal using the RIS configuration $\boldsymbol{\theta}_{l+1}$ and collect received signals in $\vect{y}_{l+1} = [\vect{y}_{l}^{\Ttran},y_{l+1}]^{\Ttran}$}
\ENDIF
 \ENDFOR 
 \STATE {Compute $\hat{\alpha}$ in \eqref{eq:amp_d_estimate}}
\STATE \textbf{Output}:   $\hat{\vect{g}} = \sqrt{\hat{\beta}_L}e^{j\hat{\omega}_L}\vect{a}(\hat{\boldsymbol{\psi}}_L)$ and $\hat{d} = \sqrt{\hat{\alpha}} e^{j\hat{\vartheta}_L}$
 \end{algorithmic}
\end{algorithm}

In the conference version  \cite{haghshenas2023efficient} of this paper, we picked $\{\boldsymbol{\theta}_1,\boldsymbol{\theta}_2\}$ randomly from $\Theta$, which was the conventional DFT codebook. 
In the next two sections, we will explain the procedure for designing a better codebook $\Theta$ and propose a novel method for initializing the first two RIS configurations that reduces the number of required pilots compared to the case where we initialize the RIS configurations randomly as in \cite{haghshenas2023efficient}.

\section{Codebook Design}
\label{sec:codebook}
In this section, we will design the codebook of RIS configurations $\Theta$ to consider during the pilot transmission to minimize the search space. To this end, we first find a set of azimuth-elevation angle pairs that form an orthogonal basis of the RIS-related channels in the far-field.
We will then explain why this codebook can also be used for pilot transmission when considering the radiative near-field, instead of designing a dedicated 
near-field codebook as in \cite{Dai2023}.

\begin{figure*}
    \centering
    \subfloat[\label{fig:Beamdemo1} ]{
    \includegraphics[width = 0.33\textwidth]{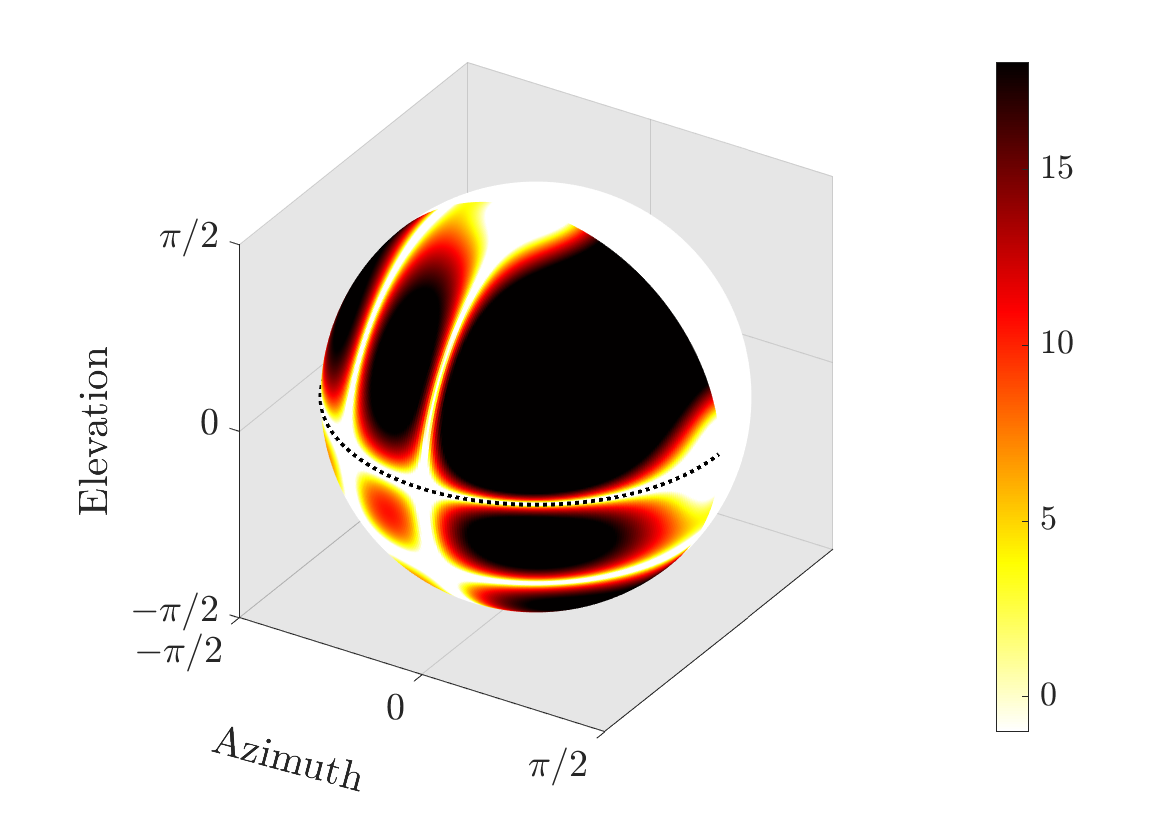}
    }
    \subfloat[\label{fig:Beamdemo2} ]{
    \includegraphics[width = 0.33\textwidth]{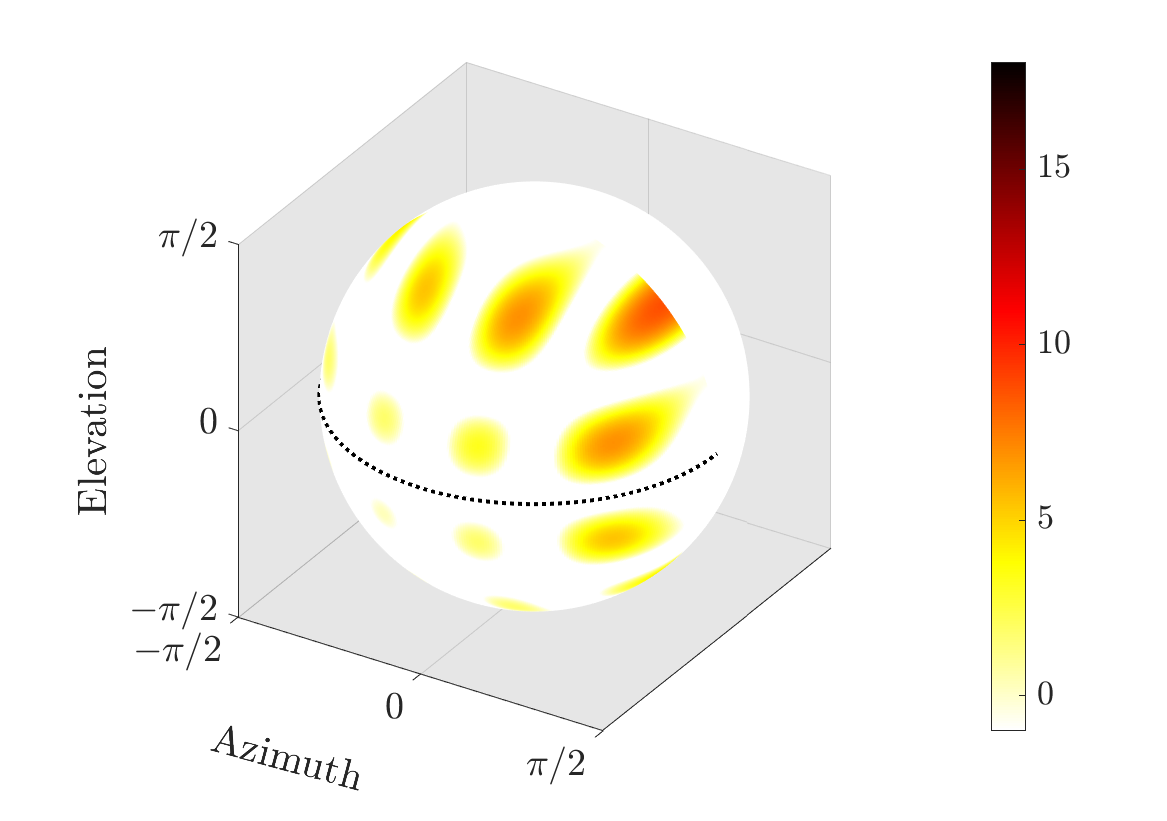}
    } 
    \subfloat[\label{fig:Beamdemo3} ]{
    \includegraphics[width = 0.33\textwidth]{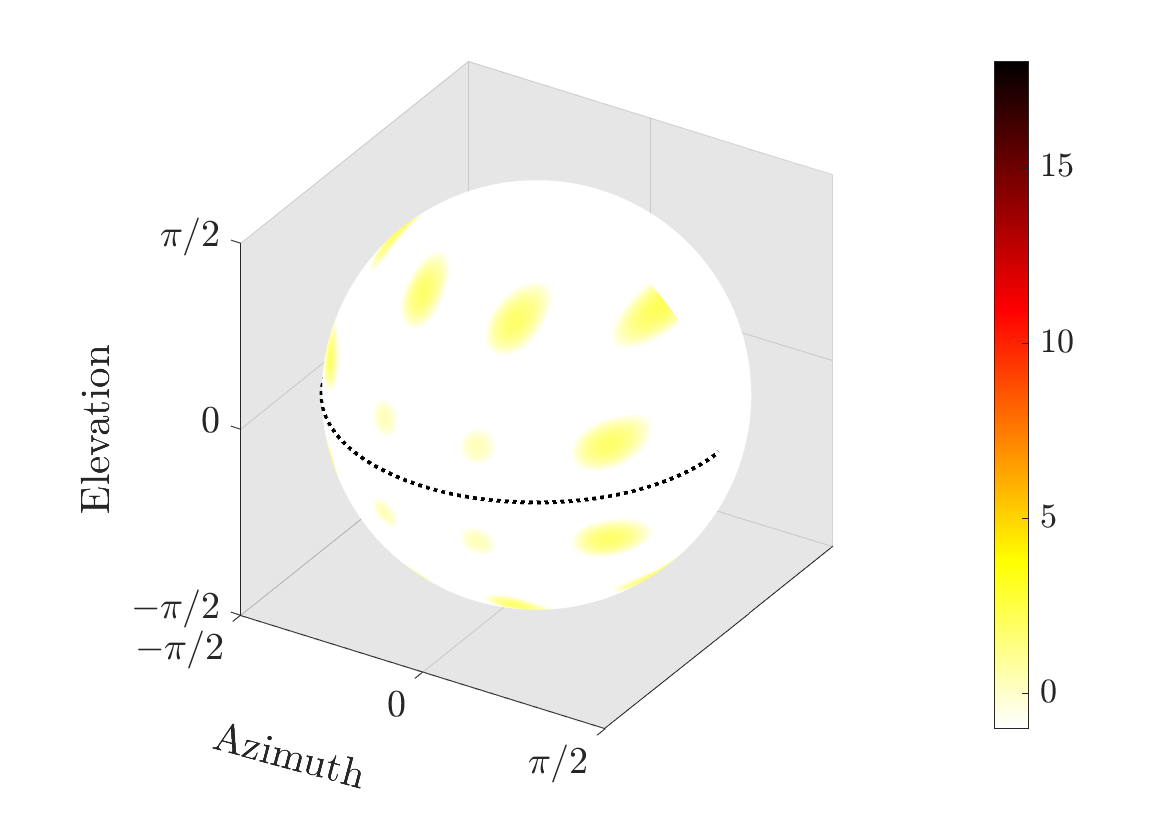}
    }
    \caption{Realization of three DFT columns in beamspace. (a) corresponds to a DFT column that is a crucial basis to construct the channel space, while (b) and (c) are two DFT columns that do not span the channel space and can be neglected.}
    \label{fig:Beamdemo}
\end{figure*}

The $N \times N$ DFT matrix is commonly used as the pilot transmission codebook when having an $N$-element RIS as it spans any $N$-dimensional vector space \cite{BZhang2020,Carvalho2020}. The DFT matrix also has special geometric properties since the columns are array response vectors of a uniform linear array that are equally spaced in wavenumbers.
If we define $\mathbf{F}_{N_\textrm{H}}\in \mathbb{C}^{N_{\textrm{H}}\times N_{\textrm{H}}}$ and $\mathbf{F}_{N_\textrm{V}} \in \mathbb{C}^{N_{\textrm{V}}\times N_{\textrm{V}}}$
as two DFT matrices such that $[\mathbf{F}_{\textrm{P}}]_{n,t} = \exp(-j2\pi(n-1)(t-1)/P)$, then the 2D DFT matrix $\mathbf{F} = \mathbf{F}_{N_{\textrm{V}}} \otimes \mathbf{F}_{N_{\textrm{H}}}$ represent equally spaced wavenumbers in the horizontal and vertical dimensions for a $N_{\textrm{H}} \times N_{\textrm{V}}$ UPA \cite{Pei2021a}. However, we will develop a codebook that is more efficient than the 2D DFT matrix when considering a UPA-structured RIS.

Let $\vect{f}_i \in \mathbb{C}^N $ be the $i$th column of the 2D DFT matrix $\vect{F}$ defined above and let $g_i(\varphi,\phi) = | \mathbf{f}_i^{\Htran} \mathbf{a}(\varphi,\phi)|^2$ denote the beamforming gain observed in the direction $\varphi \in [-\pi/2, \pi/2)$ and $\phi \in [-\pi/2, \pi/2)$ when $\vect{f}_i$ is used as the RIS phase shift vector. We want to examine if the basis vector $\mathbf{f}_i$ falls into the channel space of the RIS. To this end, Fig.~\ref{fig:Beamdemo} plots $g_i$ over the surface of a unit sphere for three different columns of the DFT matrix $\vect{F}$. The number of RIS elements is considered to be $N=64$ with $N_{\mathrm{H}}=N_{\mathrm{V}}=8$ and the element spacing is $\Delta_{\textrm{H}} = \Delta_\textrm{V} = \frac{\lambda}{4}$. It is evident that the DFT vectors realizing Fig.~\ref{fig:Beamdemo2} and Fig.~\ref{fig:Beamdemo3} do not correspond to any angular direction and hence are not needed to construct the RIS channel. On the contrary, the DFT vector that realizes Fig.~\ref{fig:Beamdemo1} serves as an important basis vector for the construction of the channel.
This implies that many columns of the 2D DFT matrix cannot provide useful information about the RIS channel and all possible RIS array response vectors can be approximately expressed using only a subset of the columns. 
There are two theoretical explanations for 
this result \cite{Hu2018a,Pizzo2020,Ozlemarraygeometry,haghshenas2023new,orfanidis2002electromagnetic}. Firstly, the wavenumbers that an impinging wave can create horizontally and vertically on a planar surface are mutually dependent, which limits the search space. Secondly, the inter-element spacing of smaller than $\lambda/2$ corresponds to spatial oversampling which increases the signal correlation between adjacent antennas and  reduces the visible region of the array.
This raises the need for a new codebook design that can efficiently span the RIS channel subspace.

To develop such a codebook, we first need to determine a set of orthogonal array response vectors that span the RIS channel subspace and then utilize them to find the RIS configurations to be used in the channel estimation phase. Using the expression of the far-field array response in \eqref{eq:array_response_farfield}, we can write the magnitude of the inner product of two array response vectors corresponding to the angle pairs $(\varphi_1,\phi_1)$ and $(\varphi_2,\phi_2)$ as 
\begin{equation} \label{eq:inner-product-arbitrary-angles}
    \begin{aligned}
        &\left |  \mathbf{a}(\varphi_2,\phi_2)^{\Htran}\mathbf{a}(\varphi_1,\phi_1) \right | 
        =  \left | \sum_{n=1}^{N} e^{j\frac{2\pi}{\lambda}  \left ( i(n)\Lambda + k(n)\Omega \right )} \right | \\
        &= \underbrace{\left |  \sum_{k=0}^{N_\textrm{V}-1} e^{j\frac{2\pi}{\lambda} \Delta_\textrm{V} k \Omega}\right |}_{S(\Omega)}\underbrace{\left |   \sum_{k^{\prime}=0}^{N_\textrm{H}-1} e^{j\frac{2\pi}{\lambda} \Delta_\textrm{H} k^{\prime}\Lambda}\right |}_{T(\Lambda)} ,
    \end{aligned}
\end{equation}
where
\begin{equation}
\label{eq:Omega}
    \Omega = \sin(\phi_2) - \sin(\phi_1),
\end{equation}
\begin{equation}
\label{eq:Lambda}
    \Lambda = \sin(\varphi_2)\cos(\phi_2) - \sin(\varphi_1)\cos(\phi_1).
\end{equation}
Using standard methods \cite{EmilMIMObook}, we can rewrite $T(\Lambda)$ and $S(\Omega)$ as
\begin{equation}
\label{eq:corr1}
    S(\Omega) = \left | \frac{\sin(\frac{\pi}{\lambda} N_\textrm{V} \Delta_\textrm{V} \Omega)}{N_\textrm{V} \sin(\frac{\pi}{\lambda} \Delta_\textrm{V} \Omega)}  \right |,
\end{equation}
\begin{equation}
\label{eq:corr2}
    T(\Lambda) = \left | \frac{\sin(\frac{\pi}{\lambda}N_\textrm{H} \Delta_\textrm{H} \Lambda)}{N_\textrm{H} \sin (\frac{\pi}{\lambda} \Delta_\textrm{H} \Lambda)} \right |.
\end{equation}
To identify the set of orthogonal array response vectors, their inner product in \eqref{eq:inner-product-arbitrary-angles} should be zero. This condition is satisfied if either $S(\Omega)$ or $T(\Lambda)$ is zero. Since $\Omega$ and $\Lambda$ are functions of two angles, we need to set an initial angle and search for other angles that satisfy the orthogonality condition. We consider $(\varphi_1,\phi_1)$ as the reference angle pair.

From \eqref{eq:corr1} we notice that $S(\frac{k\lambda}{N_\textrm{V} \Delta_\textrm{V}}) = 0$ for $ k = \pm1, \dots,\pm (N_\textrm{V}-1)$, and $S(\Omega)$ is periodic with the period of $\frac{\lambda}{\Delta_\textrm{V}}$. Considering $\phi_1$ as the reference elevation angle, we identify other elevation angles by solving $S(\Omega) = 0$. Since the function is periodic, we have $S(\Omega) = 0$ for $ k = \pm1, \dots,\pm\left \lfloor \frac{N_\textrm{V}}{2}\right \rfloor$  if 
\begin{equation}
\label{eq:finding_elevation}
    \sin(\phi_2) = \frac{k\lambda}{N_\textrm{V} \Delta_\textrm{V}} + \sin(\phi_1).
\end{equation}Solving \eqref{eq:finding_elevation} for $\phi_2$, we obtain a set of elevation angles that are mutually orthogonal. After collecting all the elevation angles from the previous step in the set $\Phi = \left  \{ \phi_1, \ldots, \phi_K \right \}$, for each $\phi_k \in \Phi$ we also need to find the azimuth angles that result in $T(\Lambda) = 0$. This is required to preserve the orthogonality among all the array response vectors with the same elevation angle. From \eqref{eq:corr2}, we observe that $T(\frac{k^\prime\lambda}{N_\textrm{H} \Delta_\textrm{H}}) = 0$ for $k^\prime = \pm 1, \dots, \pm (N_\textrm{H}-1) $, and it is periodic with period $\frac{\lambda}{\Delta_\textrm{H}}$. We consider $\varphi_1$ as the reference angle for each elevation angle $\phi_k \in \Phi,~k=1,\ldots,K,$ to identify other azimuth angles that satisfy $T(\Lambda) = 0$.  
Therefore, we construct the set
\begin{equation}
\label{eq:azSet}
    \Pi(\phi_k) = \\
    \left \{ \varphi_2: \Lambda = \cos(\phi_k)\left ( \sin(\varphi_2) - \sin(\varphi_1) \right ) = \frac{k^\prime\lambda}{N_\textrm{H} \Delta_\textrm{H}} \right \}
\end{equation}
for $k^\prime= \pm 1, \dots, \pm (N_\textrm{H}-1)$ in order to collect all the azimuth angles corresponding to the orthogonal beams associated with the elevation angle $\phi_k$. These azimuth angles are generally different for each elevation angle value. We collect all the azimuth-elevation angle pairs that are generated in this way in the set $\mathcal{B} =\left  \{ (\varphi_1,\phi_1), \dots, (\varphi_{\eta},\phi_{\eta}) \right \}$, where $|\mathcal{B}| = \eta$ specifies the dimension of the RIS channel space.
Accordingly, the codebook of RIS configurations can be obtained as 
\begin{equation}
\begin{aligned}
     &\Theta = \\&\left\{\diag \left(e^{-j\mathrm{arg}(h_1)},\ldots,e^{-j\mathrm{arg}(h_N)} \right) \vect{a}^*(\varphi,\phi) : (\varphi,\phi) \in \mathcal{B} \right \}. 
     \end{aligned}
\end{equation}

\begin{figure}
    \centering
    \includegraphics[width = \linewidth]{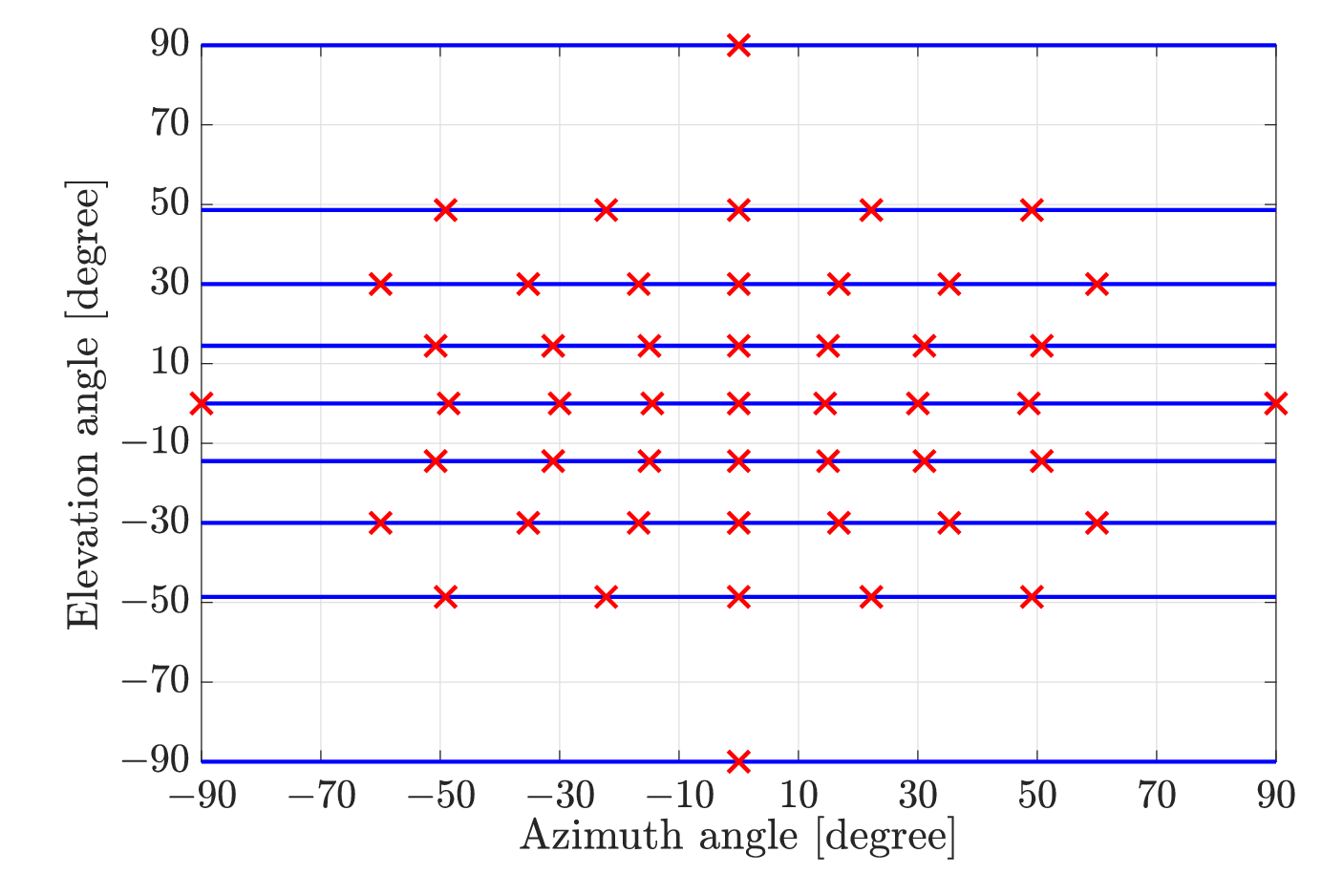}
    \caption{Orthogonal angle pairs marked with crosses for a $16 \times 16$ RIS with $\Delta_{\textrm{H}} = \Delta_{\textrm{V}} = \frac{\lambda}{4}$, computed using Algorithm~\ref{alg:Selection}.}
    \label{fig:AzEl}
\end{figure}

In summary, we have determined a set of elevation angles that generates orthogonal channel directions. This step establishes the solid blue lines in the elevation-azimuth plane in Fig.~\ref{fig:AzEl} for a $16 \times 16$ RIS with $\Delta_\textrm{H} = \Delta_\textrm{V} = \frac{\lambda}{4}$. In this figure, every two points belonging to different blue lines are mutually orthogonal because $S(\Omega)=0$. In the second step, we have specified the azimuth angles that satisfy the orthogonality condition \eqref{eq:azSet} for each blue line. In Fig.~\ref{fig:AzEl}, these angles are shown with red crosses. 
Algorithm \ref{alg:Selection} summarizes the procedure for computing this codebook. We propose to use it as input to the MLE estimation method in Algorithm \ref{Alg:ML_Estimation} to minimize the search space for RIS configurations.

\begin{small}
\begin{algorithm}[t!]
\begin{algorithmic}[1]
 \STATE \textbf{Input}: $\mathbf{h} = [h_1,\ldots,h_N]^{\Ttran}$
 \STATE Initialize $\Theta = \emptyset,\mathcal{B} = \emptyset,$ and $(\varphi_1,\phi_1)$
\STATE {Obtain the set of elevation angles as $\Phi = \left\{ \phi = \arcsin(\frac{k\lambda}{N_\textrm{V} \Delta_\textrm{V}}+\sin(\phi_1)),~k = 0, \pm1,\ldots,\pm\left \lfloor \frac{N_\textrm{V}}{2} \right \rfloor\right\}$}
 \FOR{$\phi_k \in \Phi$}  
 \FOR{$\ell = 0, \pm 1, \ldots, \pm (N_\textrm{H}-1)$}
    \STATE $\varphi \gets \arcsin\left ( \sin(\varphi_1) + \frac{\frac{\ell \lambda}{N_\textrm{H} \Delta_\textrm{H}}}{\cos(\phi_k)} \right )$ 
    \STATE $\mathcal{B} \gets \mathcal{B} \cup \{ (\varphi,\phi_k)\}$
\ENDFOR
\ENDFOR
\FOR{($\varphi_i,\phi_i) \in \mathcal{B}$}
    \STATE $\Theta \gets \Theta \cup \diag \left(e^{-j\mathrm{arg}(h_1)},\ldots,e^{-j\mathrm{arg}(h_N)} \right) \vect{a}^*(\varphi_i,\phi_i)$
\ENDFOR
 \STATE \textbf{Output}: The codebook matrix $\Theta$
\caption{The proposed codebook design.}
\label{alg:Selection}
\end{algorithmic}
\end{algorithm}
\end{small}

The codebook of RIS configurations has been designed considering a far-field scenario. Nevertheless, the same codebook can be employed for estimating the channel when the user is in the near-field region of the RIS. We will demonstrate this numerically in Section \ref{sec:numericalRes}, while the theoretical reason is clarified in the following remark.

\begin{remark}
  Based on the Weyl identity \cite{Pizzo2020}, \cite[Chapter 2, eq.~2.2.27]{chew1995waves}, any spherical wave can be expressed as a summation of an infinite number of plane waves. Therefore, a near-field channel can be represented as a summation of far-field channels which implies that the same basis vectors that span any far-field channel can be utilized to construct any near-field channel. Consequently, the codebook of RIS configurations, designed based on array responses of far-field channels, can be utilized during pilot transmission to excite all near-field channel dimensions. We stress that the final LOS channel estimate is not restricted to come from this codebook.
\end{remark}

\begin{figure}
    \centering   \includegraphics[width = \linewidth]{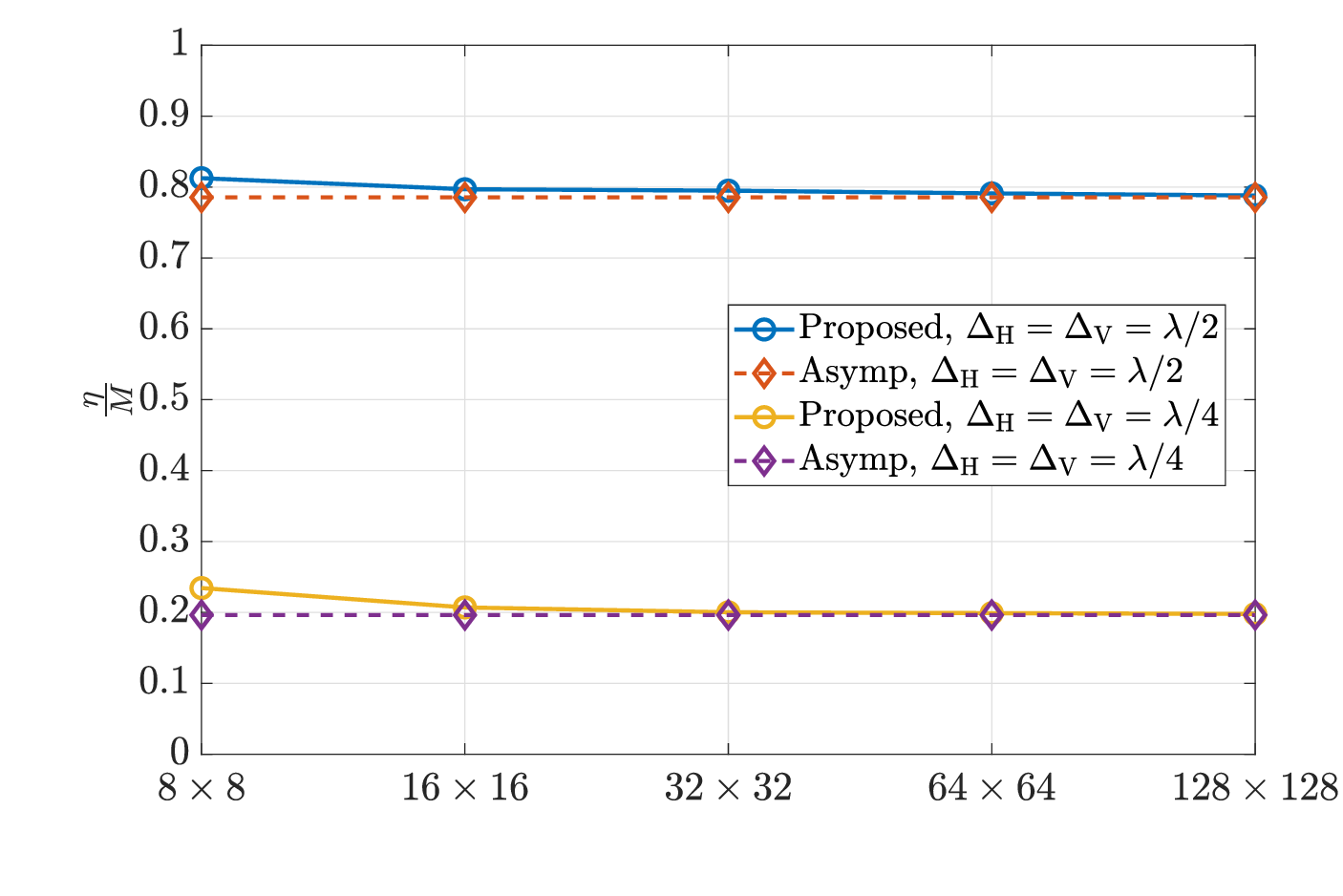}
    \caption{The DOF approximation in \eqref{eq:DOF} is compared with the exact size of the proposed codebook in Algorithm~\ref{alg:Selection}.} 
    \label{fig:approx}
\end{figure}

The dimension $\eta = |\mathcal{B}|$ of the codebook can be predicted using asymptotic results from \cite{Hu2018a,Pizzo2020}, which says that the spatial degrees-of-freedom (DOF) of a continuous UPA with area $N_\textrm{H} \Delta_\textrm{H} N_\textrm{V} \Delta_\textrm{V}$ is
\begin{equation}
\label{eq:DOF}
     \eta \approx \frac{\pi}{\lambda^2} N_\textrm{H} \Delta_\textrm{H} N_\textrm{V} \Delta_\textrm{V}.
\end{equation}
To examine under which conditions the above approximation is tight, we present in Fig.~\ref{fig:approx}, the ratio between $\eta$ and the number of RIS elements (i.e., $\eta/N$) for different numbers of reflecting elements and inter-element spacing. The dashed lines correspond to the DOF approximation in the right-hand side of \eqref{eq:DOF}, while the solid lines represent the size of the codebook generated by Algorithm \ref{alg:Selection}. It is evident that the approximation is tight for large arrays, while it underestimates the number of required codebook entries for smaller arrays.

\begin{figure}[t!]
\centering 
	\begin{overpic}[width=.8\columnwidth]{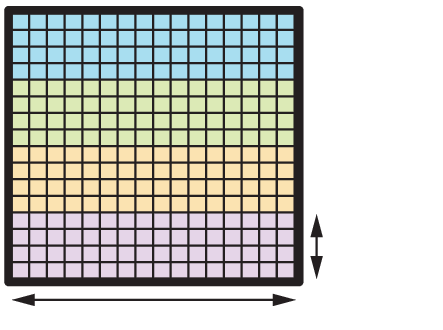}
	 \put (21,1) {$N_{\mathrm{H}}$ columns}
	 \put (75,20) {Sub-RIS with}
	 \put (75,15) {$2^{x-1}$ rows}
\end{overpic}  
    \caption{A RIS with $N_{\mathrm{H}} \times N_{\mathrm{H}}$ elements is divided into $N_{\mathrm{H}}/2^{x-1}$ sub-RISs each with $2^{x-1}$ rows and $N_{\mathrm{H}}$ columns to generate a wide beam. In this example, $N_{\mathrm{H}}=16$ and $x=3$.}
    \label{fig:sub_RISs}
\end{figure}
\section{Initial Beam Design}
\label{sec:Init}

Algorithm \ref{Alg:ML_Estimation} can be initialized by considering two random columns from the codebook matrix $\Theta$, which are then removed from it, as suggested in our conference paper \cite{haghshenas2023efficient}. This corresponds to reflecting the pilot signals into two independently selected narrow beam directions. The probability that these beams match with the user location is only $2/|\mathcal{B}|$ if there is no prior information about the user location. It is only when the MLE algorithm has a hunch of where the user is that it quickly converges by ruling out spatial ambiguities, and there is a risk that a random initialization will not provide that information.
In this section, we propose an alternative way of initializing the RIS configuration in Algorithm \ref{Alg:ML_Estimation} by using two wider beams that can simultaneously explore multiple angular directions and guarantee useful information from the first pilot transmissions irrespective of the user location.

\subsection{Wide Beam Design}
\label{subsec:widebeam-design}

The beamwidth of the signal emitted from an array is inversely proportional to the array's aperture length.
To produce wider beams, the RIS can be divided into multiple sections, each having a smaller aperture and therefore generating a wider beam. Each section can then be configured to beamform in a different angular direction so that the superposition of the radiated signal remains wide. We will now propose a careful wide-beam design.

\begin{figure}[t!]
    \centering
    \includegraphics[width = 0.9\linewidth]{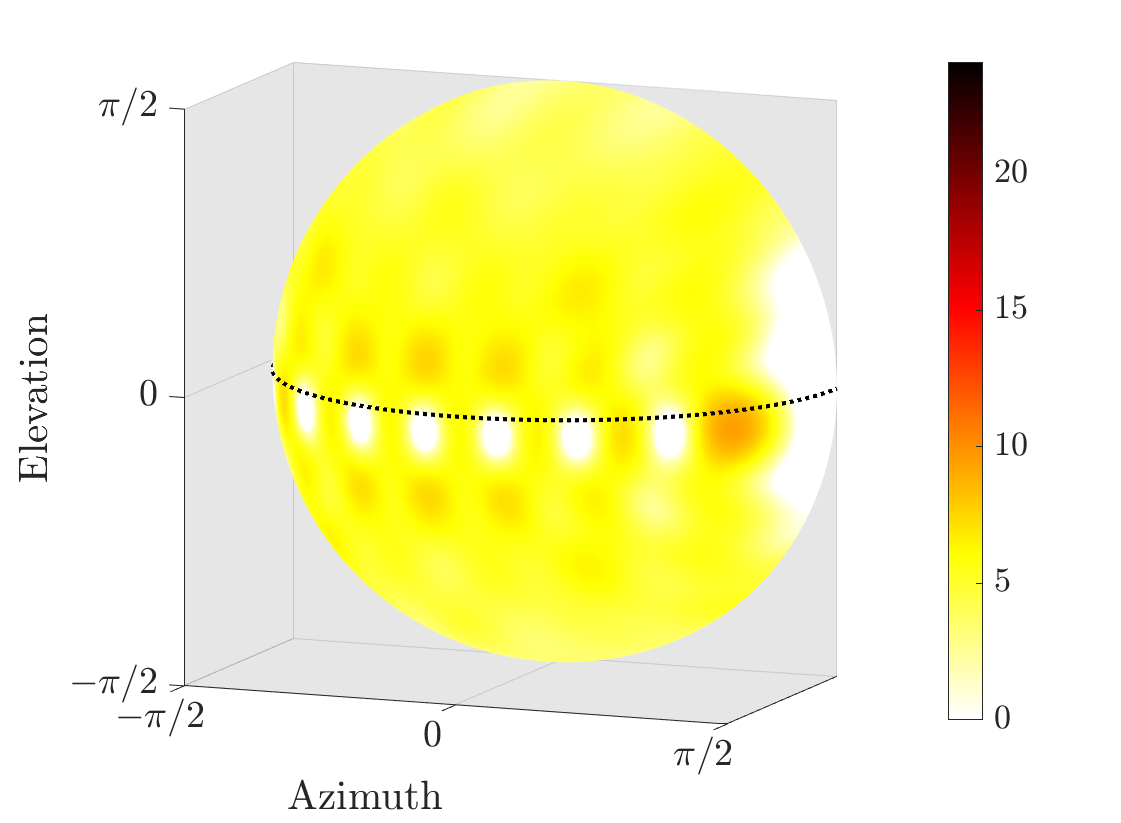}
    \caption{The nearly isotropic beam generated by a RIS with $N_{\mathrm{H}} = N_{\mathrm{V}} = 16$ and $\Delta_{\mathrm{H}} = \Delta_{\mathrm{V}} = \lambda/4$. For this specific setup, we have $N_S = 32$ and $S = 8$.}
    \label{fig:Wideshape}
\end{figure}

We divide the RIS into $S$ equal-sized sub-RIS sections, each having $N_S = N/S$ elements. We let $\eta_s$ be the dimension of the channel space for each sub-RIS, where $\eta_s$ can be obtained using Algorithm \ref{alg:Selection}.
We can beamform simultaneously in all directions if we divide the RIS in a way that $\eta_s \approx S$ and let each sub-RIS point a beam in the direction specified by one of the $\eta_s$ angle pairs. To find the number of elements in each sub-RIS, we use the DOF approximation in 
\eqref{eq:DOF} and set 
\begin{equation}
\underbrace{\frac{\pi}{\lambda^2}N_S\Delta_{\mathrm{H}}\Delta_{\mathrm{V}}}_{\eta_s} = \underbrace{\frac{N}{N_S}}_{S}, 
\end{equation}
which results in 
    $N_S = \sqrt{\frac{N_{\mathrm{H}}N_{\mathrm{V}}\lambda^2}{\pi\Delta_{\mathrm{H}}\Delta_\mathrm{V}}}$.
For a square-shaped RIS with $N_{\mathrm{H}} = N_{\mathrm{V}}$ and $\Delta_{\mathrm{H}} = \Delta_{\mathrm{V}} = \lambda/2^x,~x \in \mathbb{N}$, we obtain 
\begin{equation}
    N_S = \frac{N_{\mathrm{H}}\lambda}{\Delta_{\mathrm{H}}}\sqrt{\frac{1}{\pi}} \overset{(a)}{\approx} \frac{N_{\mathrm{H}}\lambda}{2\Delta_{\mathrm{H}}} = 2^{x-1}N_{\mathrm{H}},
\end{equation}
where we have used $\sqrt{1/\pi} \approx 1/2$ in (a). Thus, the number of sub-RISs will be given by $S = N_{\mathrm{H}}/2^{x-1}$. There are several ways to split RIS into $S$ sub-RISs of $N_S$ elements. Here, we split the RIS into $S$ sub-RIS each having $2^{x-1}$ rows and $N_{\mathrm{H}}$ columns, as depicted in Fig.~\ref{fig:sub_RISs}. 
The set of orthogonal angle-pairs $\mathcal{B}_s$ can  be obtained from Algorithm \ref{alg:Selection} for a $2^{x-1} \times N_{\mathrm{H}}$ sub-RIS, where $|\mathcal{B}_s| = \eta_s$. The RIS beamforming vector
\begin{equation}
\label{eq:wideb}
    \begin{aligned}
    \boldsymbol{w} = \Bigl[ \bigl[\mathbf{a}(\varphi_1,\phi_1)\bigr]^{\Ttran}_{1:N_\textrm{s}}, \bigl[\mathbf{a}(\varphi_2,\phi_2)\bigr]^{\Ttran}_{N_\textrm{s}+1:2N_\textrm{s}},\ldots, \\
    \bigl[\mathbf{a}(\varphi_S,\phi_S)\bigr]^{\Ttran}_{(S-1)N_\textrm{s}+1:N}
    \Bigr]
    \end{aligned}
\end{equation}
produces an approximately isotropic beam that covers all the angular directions, where $(\varphi_s,\phi_s) \in \mathcal{B}_s,~s = 1,2,\ldots,S$. This is illustrated in Fig.~\ref{fig:Wideshape} for a RIS with $N_\mathrm{H} = N_{\mathrm{V}} = 16$ and $\Delta_{\mathrm{H}} = \Delta_{\mathrm{V}} = \lambda/4$. The beam generated by the RIS configuration in $\eqref{eq:wideb}$ is a wide beam that covers all directions.

The proposed scheme ensures that all the sub-RISs use mutually orthogonal beamforming vectors. The fact that the beams do not interfere with one another has two key consequences. Firstly, each beam maintains its wider beamwidth since there is no constructive interference between neighboring beams. Secondly, the combination of $S$ beams is an even wider beampattern that has a nearly constant gain over all angles. the fact that we generate angle pairs based on sub-RISs with $N_S$ elements and not the full-aperture RIS is essential. Our approach is thereby different from the prior works in \cite{Rui2020,Hierar2016}, where the beams are generated considering the full-aperture RIS. In these works, the beams are distributed over the sub-RISs based on the beam distances to reduce inter-beam interference, while our method intrinsically guarantees zero inter-beam interference, as will be shown in the following. 

 \begin{figure*}[t!]
    \centering
    \subfloat[\label{fig:sideExp1} ]{
    \includegraphics[width = 0.4\textwidth]{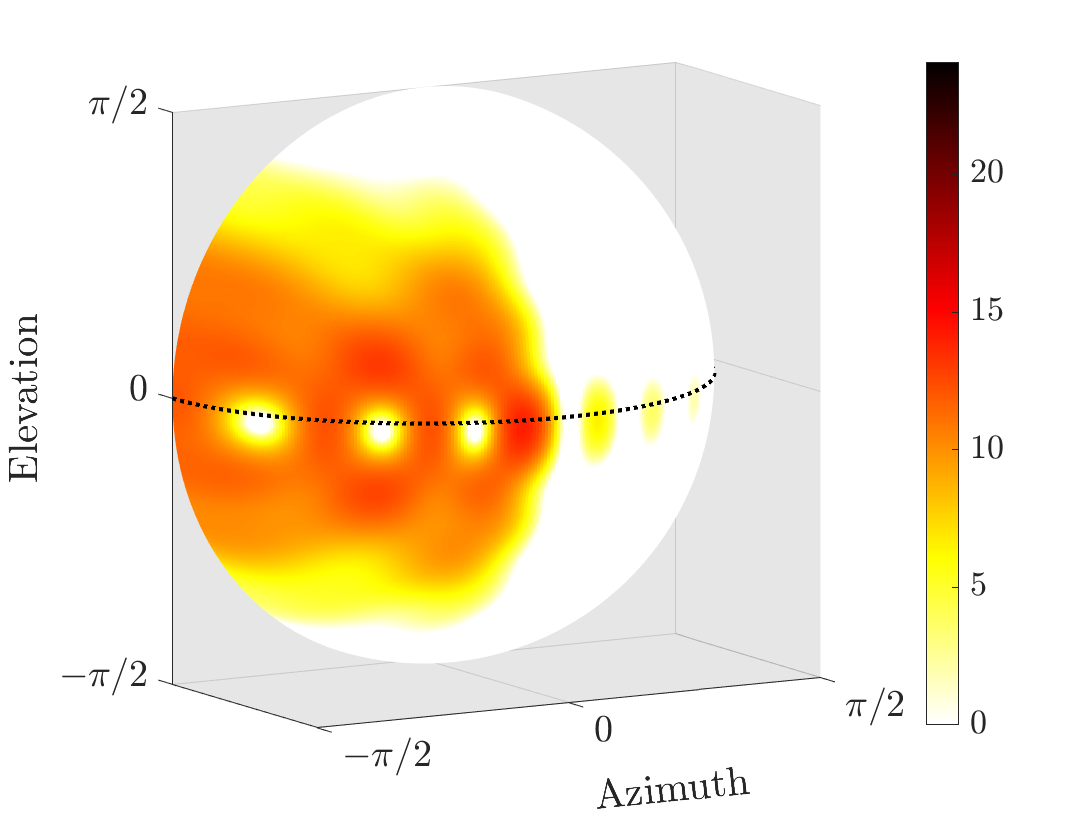}
    }
    \subfloat[\label{fig:sideExp2} ]{
    \includegraphics[width = 0.4\textwidth]{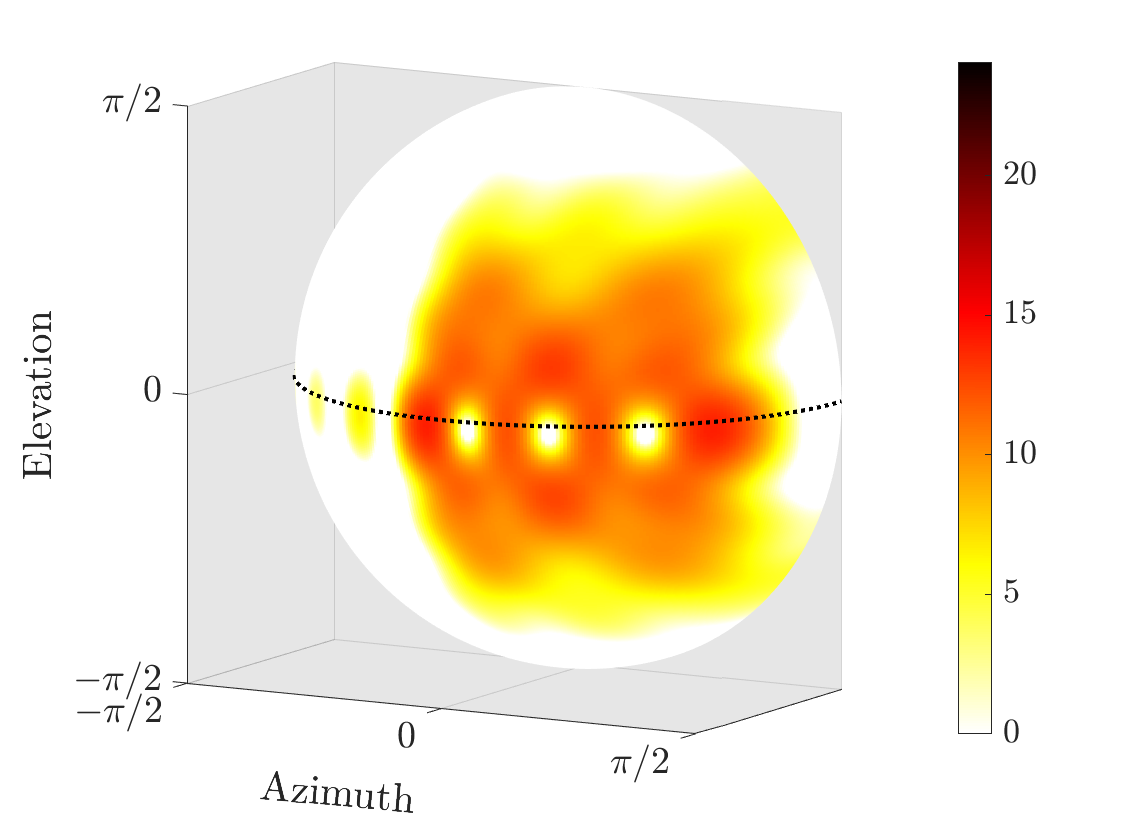}
    }
    \caption{Two wide beams generated by a RIS with $N_{\mathrm{H}} = N_{\mathrm{V}}  = 16$ and $\Delta_{\mathrm{H}} = \Delta_{\mathrm{V}} = \lambda/4$, where each beam is designed to cover half of the angles seen by the RIS. Such wide beam design assures that the two initial pilots in  Algorithm \ref{Alg:ML_Estimation} cover all potential user angles, and the two received signals can be compared to obtain a first estimate of the user channel.}
    \label{fig:sideExp}
\end{figure*}

 Let $\vect{b}_1(\varphi,\phi)$ be the array response vector of the first sub-RIS for a wave coming from the angle $(\varphi,\phi)$. The array response vector of the $s$th sub-RIS will be given by $\vect{b}_s (\varphi,\phi)= e^{j\frac{2\pi}{\lambda}\Delta_{\mathrm{V}}(2^{x-1})(s-1)\sin(\phi)} \vect{b}_1(\varphi,\phi)$. If we assign two different angle pairs $\{(\varphi_1,\phi_1),(\varphi_2,\phi_2)\} \in \mathcal{B}_s$ to two different sub-RISs with indices $s_1$ and $s_2$, we have 
 \begin{equation}
\begin{aligned}
    &|\mathbf{b}_{s_1}(\varphi_1,\phi_1)^\H \mathbf{b}_{s_2}(\varphi_2,\phi_2)| = \\ &
   \Big|e^{j\frac{2\pi}{\lambda}\Delta_{\mathrm{V}}(2^{x-1})\big( (s_1-1) \sin(\phi_1) - (s_2 -1)\sin(\phi_2)\big)}\Big| \, \times \\  &\Big|\mathbf{b}_1(\varphi_1,\phi_1)^\H \mathbf{b}_1(\varphi_2,\phi_2)\Big| = \\ & \underbrace{\left |  \sum_{k=0}^{2^{x-1}-1} e^{j\frac{2\pi}{\lambda} \Delta_\textrm{V} k \Omega}\right |}_{S(\Omega)}\underbrace{\left |   \sum_{k^{\prime}=0}^{N_\textrm{H}-1} e^{j\frac{2\pi}{\lambda} \Delta_\textrm{H} k^{\prime}\Lambda}\right |}_{T(\Lambda)} = 0,
\end{aligned}
\end{equation}where $\Omega$, $\Lambda$, $S(\Omega)$, and $T(\Lambda)$ are given in \eqref{eq:Omega}, \eqref{eq:Lambda}, \eqref{eq:corr1}, and \eqref{eq:corr2}, respectively. 

We can follow a similar methodology to design two wide beams that each cover half of the angular directions. This is what we need for the initialization of Algorithm \ref{Alg:ML_Estimation}. To this end, we split the set of orthogonal angle pairs in $\mathcal{B}_s$ into two subsets $\mathcal{B}_{s,1}$ and $\mathcal{B}_{s,2}$ such that $\mathcal{B}_{s,1} \cap \mathcal{B}_{s,2} = \emptyset$, $|\mathcal{B}_{s,1}|  = \lfloor \eta_s/2 \rfloor$, and $|\mathcal{B}_{s,2}| = \lceil \eta_s/2 \rceil$. We then construct two RIS beamforming vectors based on the angle pairs present in these two subsets as 
\begin{equation}
\label{eq:wideb2}
    \begin{aligned}
    \boldsymbol{w}_i = \Bigl[ \bigl[\mathbf{a}(\varphi_1,\phi_1)\bigr]^{\Ttran}_{1:2N_\textrm{s}}, \bigl[\mathbf{a}(\varphi_2,\phi_2)\bigr]^{\Ttran}_{2N_\textrm{s}+1:4N_\textrm{s}},\ldots, \\
    \bigl[\mathbf{a}(\varphi_S,\phi_S)\bigr]^{\Ttran}_{(S-1)2N_\textrm{s}+1:N}
    \Bigr],~i=1,2.
    \end{aligned}
\end{equation}In $\boldsymbol{w}_i$, each two sub-RISs beamform towards one of the angle pairs in $\mathcal{B}_{s,i}$. The set of $\eta_s$ orthogonal angle pairs can be split between the $\mathcal{B}_{s,1}$ and $\mathcal{B}_{s,2}$ in a way that each configuration vector $\boldsymbol{\theta}_i$ produces a beam covering one side (either left or right) of the RIS. This scenario is portrayed in Fig.~\ref{fig:sideExp}. 

We can now utilize the beamforming vectors in \eqref{eq:wideb2} to establish two initial RIS configurations as 
\begin{equation}
\label{eq:initial_RIS}
    \boldsymbol{\theta}_i = \diag \left(e^{-j\mathrm{arg}(h_1)},\ldots,e^{-j\mathrm{arg}(h_N)} \right) \boldsymbol{w}^*_i.
\end{equation}Using these two new configurations as the initial RIS phase shift vectors in Algorithm \ref{Alg:ML_Estimation}, we can ensure that the user is always covered by one of the beams reflected by the RIS during the first two pilot transmissions, while the two received signals can be compared to obtain the first MLE. We refer to the configurations in \eqref{eq:initial_RIS} as wide beam configurations.

\section{Extension to Multi-Antenna BS}
The proposed MLE framework can be extended to accommodate scenarios with multiple antennas at the BS. In this section, we briefly explain the parametric MLE approach and adaptive RIS configuration strategy for such scenario. 

Assume that the BS is equipped with $M$ antennas. The channel between the BS and the RIS and that between the BS and the user are respectively denoted as $\vect{H} \in \mathbb{C}^{M \times N}$ and $\vect{d} \in \mathbb{C}^M$. As previously stated, the channel between the BS and the RIS is assumed to be known in advance as the locations of the RIS and BS are fixed.

The received signal at the BS at the $l$th pilot instance is given by 
\begin{equation}
\begin{aligned}
\label{eq:received-pilot-multiantenna}
    \vect{y}_l &= \left(\vect{H}\diag(\vect{g})\boldsymbol{\theta}_l+\vect{d}\right)x_l + \vect{w}_l \\ & = \left( \vect{H} \odot (\vect{1}_M \kron \boldsymbol{\theta}_l^\T)\vect{g} + \vect{d}\right)x_l + \vect{w}_l,
    \end{aligned}
\end{equation}where
$\vect{w}_l \sim \CN(0,\sigma^2 \vect{I}_M)$ represents the independent complex Gaussian receiver
noise.

Recall that the user transmits the deterministic pilot signal  $x_l = \sqrt{P}_{\mathrm{p}}$; the collected received signal at the BS can then be represented as
\begin{equation}
    \vect{y} = \Biggl( \underbrace{\Bigl( (\boldsymbol{1}_{L} \kron \vect{H}) \odot (\vect{B} \kron \boldsymbol{1}_M)\Bigr)}_{ = \vect{F} \in \mathbb{C}^{LM \times N} }\vect{g} + (\vect{1}_L \kron \vect{d}) \Biggr) \sqrt{P_\mathrm{p}}+ \bar{\vect{w}},
\end{equation}
where $\vect{y} = [\vect{y}_1^\T, \vect{y}_2^\T\ldots \vect{y}_L^\T]^\T \in \mathbb{C}^{LM}$ and $\bar{\vect{w}} = [\vect{w}_1^\T,\vect{w}_2^\T\ldots\vect{w}_L^\T]^\T \in \mathbb{C}^{LM}$.

Here, we briefly explain the MLE approach for the multi-antenna BS scenario. As the derivations closely mirror those of the single-antenna BS case, we refrain from delving into details. The MLE problem for estimating $\vect{g}$ and $\vect{d}$ is formulated as follows:
\begin{align}
\label{eq:g-d-multiple-antenna-estimate}
    \{\hat{\vect{g}},\hat{d}\} & =  \argmin{\vect{g} \in \mathcal{A},\vect{d}\in \mathbb{C}^M}~ \left\|\vect{y} - \Bigl(\vect{F} \vect{g} + (\vect{1}\kron \vect{d}) \Bigr) \sqrt{P_{\mathrm{p}}} \right\|^2 \notag \\ 
    & = \argmin{\vect{g} \in \mathcal{A},\vect{d}\in \mathbb{C}^M} P_{\mathrm{p}}\left\| \vect{F} \vect{g}\right \|^2 + P_{\mathrm{p}}L\left\| \vect{d} \right\|^2 - 2\sqrt{P_{\mathrm{p}}}\mathrm{Re}\left \{ \vect{y}^\H \vect{Fg}\right \} \notag\\
    & - 2\sqrt{P_{\mathrm{p}}} \mathrm{Re} \left \{ (\vect{y} - \sqrt{P_{\mathrm{p}}}\vect{F}\vect{g})^\H(\vect{1} \kron \vect{d})\right\}.
\end{align}
Similar to the single-antenna BS case, the estimation process starts by determining the direct channel as
\begin{equation}
    \hat{\vect{d}} = \argmin{\vect{d}\in \mathbb{C}^M}\, P_{\mathrm{p}}L\left\| \vect{d} \right\|^2 - 2\sqrt{P_\mathrm{p}} \mathrm{Re} \left \{ (\vect{y} - \sqrt{P_\mathrm{p}}\vect{F}\vect{g})^\H(\vect{1} \kron \vect{d})\right\}.
\end{equation}
The objective function tunrs out to be quadratic with respect to $\vect{d}$, with its minimum being given by:
\begin{equation}
\label{eq:d-vect-estimated}
    \hat{\vect{d}} = \frac{1}{\sqrt{P_\mathrm{p}}L}\left( \boldsymbol{1}_L^\T \kron \boldsymbol{I}_M\right)\left( \vect{y} - \sqrt{P_{\mathrm{p}}}\vect{F}\vect{g}\right). 
\end{equation}
Substituting \eqref{eq:d-vect-estimated} into \eqref{eq:g-d-multiple-antenna-estimate} and utilizing the channel structure in \eqref{eq:A-set}, the MLE problem for estimating $\vect{g}$ is expressed as 
\begin{align}
\label{eq:g-estimate-breakdown}
     & \{ \hat{\beta}, \hat{\omega}, \hat{\boldsymbol{\psi}} \} = \notag \\ 
     & \argmin{\substack{\beta \geq 0,\omega \in [0,2\pi),\\ \boldsymbol{\psi} \in \Psi}}  P_\mathrm{p}\beta \left( \| \vect{F}\vect{a}(\boldsymbol{\psi})\|^2  - \frac{1}{L}  \| \left( \boldsymbol{1}_L^\T \kron \boldsymbol{I}_M\right)\vect{F}\vect{a}(\boldsymbol{\psi}) \|^2 \right) \notag \\ 
     & - 2\sqrt{P_\mathrm{p}\beta}\mathrm{Re}\left\{ e^{j\omega}\vect{y}^\H\left( \boldsymbol{I}_{ML} - \frac{1}{L}\left( \boldsymbol{1}_{L\times L}\kron \boldsymbol{I}_{M}\right)  \right)\vect{F}\vect{a}(\boldsymbol{\psi})\right\}.
\end{align}
The phase $\omega$ exclusively appears in the second line of \eqref{eq:g-estimate-breakdown} and its optimal value is obtained as 
\begin{align}
\label{eq:omega-estimate-multiple-antenna}
    \hat{\omega} = -\arg\Biggl( \vect{y}^\H\left( \boldsymbol{I}_{ML} - \frac{1}{L}\left( \boldsymbol{1}_{L\times L}\kron \boldsymbol{I}_{M}\right)  \right)\vect{F}\vect{a}(\boldsymbol{\psi}) \Biggr).
\end{align}
Substituting \eqref{eq:omega-estimate-multiple-antenna} into \eqref{eq:g-estimate-breakdown}, we obtain:
\begin{align}
\label{eq:g-without-phase}
    & \{ \hat{\beta}, \hat{\boldsymbol{\psi}} \} = \notag \\
    & \argmin{\beta \geq 0,  \boldsymbol{\psi} \in \Psi} \, P_\mathrm{p}\beta \left( \| \vect{F}\vect{a}(\boldsymbol{\psi})\|^2  - \frac{1}{L}  \| \left( \boldsymbol{1}_L^\T \kron \boldsymbol{I}_M\right)\vect{F}\vect{a}(\boldsymbol{\psi}) \|^2 \right) \notag \\ 
    & - 2\sqrt{P_\mathrm{p}\beta}\Biggl| \vect{y}^\H\left( \boldsymbol{I}_{ML} - \frac{1}{L}\left( \boldsymbol{1}_{L\times L}\kron \boldsymbol{I}_{M}\right)  \right)\vect{F}\vect{a}(\boldsymbol{\psi}) \Biggr|. 
\end{align}
This objective function is a quadratic function of $\sqrt{\beta}$, with its minimum attained at
\begin{align}
\label{eq:beta-estimate-multipleantennas}
    \sqrt{\hat{\beta}} = \frac{1}{\sqrt{P_\mathrm{p}}}\frac{\Biggl| \vect{y}^\H\left( \boldsymbol{I}_{ML} - \frac{1}{L}\left( \boldsymbol{1}_{L\times L}\kron \boldsymbol{I}_{M}\right)  \right)\vect{F}\vect{a}(\boldsymbol{\psi}) \Biggr|}{\| \vect{F}\vect{a}(\boldsymbol{\psi})\|^2  - \frac{1}{L}  \| \left( \boldsymbol{1}_L^\T \kron \boldsymbol{I}_M\right)\vect{F}\vect{a}(\boldsymbol{\psi}) \|^2}.
\end{align}
Plugging \eqref{eq:beta-estimate-multipleantennas} into \eqref{eq:g-without-phase}, we arrive at  
\begin{align}
      \hat{\boldsymbol{\psi}} = \argmax{\boldsymbol{\psi} \in \Psi}\, \, \frac{\Biggl| \vect{y}^\H\left( \boldsymbol{I}_{ML} - \frac{1}{L}\left( \boldsymbol{1}_{L\times L}\kron \boldsymbol{I}_{M}\right)  \right)\vect{F}\vect{a}(\boldsymbol{\psi}) \Biggr|^2}{\| \vect{F}\vect{a}(\boldsymbol{\psi})\|^2  - \frac{1}{L}  \| \left( \boldsymbol{1}_L^\T \kron \boldsymbol{I}_M\right)\vect{F}\vect{a}(\boldsymbol{\psi}) \|^2}.
\end{align}
Similarly to \eqref{eq:psi-estimate}, $\hat{\boldsymbol{\psi}}$ is determined through a grid search. The following proposition summarizes the proposed MLE for the multi-antenna BS scenario.  

\begin{proposition}
    \label{th:main-result-multiantenna}
   The parametric MLEs  $\hat{\vect{g}} = \sqrt{\hat{\beta}} e^{j\hat{\omega}}\vect{a}(\hat{\boldsymbol{\psi}})$ and $\hat{\vect{d}}$ when multiple antennas are employed at the BS are given by 
\vspace{-0.5mm}
 \begin{align}
 \label{eq:AoA_estimate_multiantenna} \hat{\boldsymbol{\psi}} &= \argmax{\boldsymbol{\psi} \in \Psi }\, \, \frac{\Biggl| \vect{y}^{\Htran}\left( \boldsymbol{I}_{ML} - \frac{1}{L}\left( \boldsymbol{1}_{L\times L}\kron \boldsymbol{I}_{M}\right)  \right)\vect{F}\vect{a}(\boldsymbol{\psi}) \Biggr|^2}{\| \vect{F}\vect{a}(\boldsymbol{\psi})\|^2  - \frac{1}{L}  \| \left( \boldsymbol{1}_L^\T \kron \boldsymbol{I}_M\right)\vect{F}\vect{a}(\boldsymbol{\psi}) \|^2} \\
 \label{eq:phase_g_estimate_multiantenna}\hat{\omega} &= -\arg\Biggl( \vect{y}^{\Htran}\left( \boldsymbol{I}_{ML} - \frac{1}{L}\left( \boldsymbol{1}_{L\times L}\kron \boldsymbol{I}_{M}\right)  \right)\vect{F}\vect{a}(\boldsymbol{\psi}) \Biggr), \\
   \label{eq:amp_g_estimate-multiantenna}\hat{\beta}&= \frac{1}{P_{\mathrm{p}}}\left(\frac{\Biggl| \vect{y}^{\Htran} \left( \boldsymbol{I}_{ML} - \frac{1}{L}\left( \boldsymbol{1}_{L\times L}\kron \boldsymbol{I}_{M}\right)  \right)\vect{F}\vect{a}(\boldsymbol{\psi}) \Biggr|}{\| \vect{F}\vect{a}(\boldsymbol{\psi})\|^2  - \frac{1}{L}  \| \left( \boldsymbol{1}_L^\T \kron \boldsymbol{I}_M\right)\vect{F}\vect{a}(\boldsymbol{\psi}) \|^2} \right)^2,\\    
     \label{eq:d_estimate_multiantenna}
     \hat{\vect{d}}&= \frac{1}{\sqrt{P_\mathrm{p}}L}\left( \boldsymbol{1}_L^{\Ttran} \kron \boldsymbol{I}_M\right)\left( \vect{y} - \sqrt{P_\mathrm{p}}\vect{F}\hat{\vect{g}}\right).
 \end{align}
\end{proposition}

As in the single-antenna case, we utilize adaptive RIS configuration to gradually refine the estimates. From \eqref{eq:received-pilot-multiantenna}, the SE-maximizing RIS configuration, given the outcome of the channel estimation after $l$ pilot transmissions $\hat{\vect{g}}_l$ and $\hat{\vect{d}}_l$, is obtained by solving the following problem 
\begin{equation}
\label{eq:SE-maximizing_theta}
\bar{\boldsymbol{\theta}}_l = \argmax{\boldsymbol{\theta}}\,\|\hat{\vect{G}}_l\boldsymbol{\theta} + \hat{\vect{d}}_l\|^2,~ \mathrm{s.t.}\,\, |\theta_n| = 1, \,n = 1,\ldots, N, 
\end{equation}where $\hat{\vect{G}}_l = \vect{H}\diag(\hat{\vect{g}_l})$. This problem is not convex; we thus target a sub-optimal solution by alternate optimization of the RIS phase shifts, where we iteratively optimize one of the phase shifts having other fixed. Using this approach, the optimal phase shift for the $n$th element is obtained as \cite{Ramezani2021Backscatter}
\begin{equation}
\label{eq:optimal_theta_nhat}
    \bar{\theta}_{\hat{n}} = e^{-j\arg\left(\sum_{n\neq \hat{n}} \theta^*_n [\Tilde{\vect{G}}_l]_{n,\hat{n}} + \sum_m[\hat{\vect{d}}_l]_m^* [\hat{\vect{G}}_l]_{m,\hat{n}}\right)}.
\end{equation} where $\Tilde{\vect{G}}_l = \hat{\vect{G}}_l^\H \hat{\vect{G}}_l$ and $[\cdot]_{m,n}$ represents the entry in the $m$th row and $n$th column.
We alternately optimize the phase shifts in an iterative manner until convergence is achieved. The next RIS configuration for pilot transmission is then picked from the set of unused configurations in the codebook $\Theta$ based on \eqref{eq:new_config}.    

\section{Numerical Results}
\label{sec:numericalRes}

In this section, we evaluate the performance of the proposed algorithm, pilot codebook, and initialization through numerical simulations. 
We first examine the advantage of the wide beam configurations to initialize the MLE algorithm. Afterward, we evaluate the convergence performance of the algorithm to achieve the channel capacity in both far-field and near-field scenarios.  Lastly, we assess the capability of the proposed scheme in tracking channel variations.  

\subsection{Simulation Setup}
Unless otherwise specified, we use the following setup throughout the simulations. We consider a RIS-aided communication scenario at the carrier frequency of $28\,$GHz, where the RIS is equipped with $1024$ elements. Specifically, we assume that $N_{\mathrm{H}}=N_{\mathrm{V}} = 32$ and $\Delta_{\mathrm{H}}=\Delta_{\mathrm{V}} = \frac{\lambda}{2}$. Similarly to \cite{ramezani2022bookchapter}, we consider $d_{\mathrm{FA}}/10$ as the border between near-field and far-field regions of the RIS where $d_{\mathrm{FA}}$ denotes the Fraunhofer array distance that is computed as $d_{\mathrm{FA}} = 2W^2/\lambda$ with $W = \sqrt{(N_{\mathrm{H}}\Delta_{\mathrm{H}})^2 + (N_{\mathrm{V}}\Delta_{\mathrm{V}})^2}$ being the diagonal of the RIS. With the setup described above, we have $d_{\mathrm{FA}} \approx 11\,$m; thus, any user located closer than $1.1\,$m from the RIS is in its radiative near-field region. We assume that the direct path between the user and the BS is NLOS and $64$ times stronger than the cascaded path via a single RIS element, i.e., $d \sim \mathcal{N}_{\mathbb{C}}(0,64|h_ng_n|^2)$. We further define the per-element SNR during the data transmission as
\begin{equation}
\mathrm{SNR}_{\mathrm{d}} = \frac{P_{\mathrm{d}}}{\sigma^2} | h_n g_n |^2,
\end{equation}
and assume that the pilot SNR is $10\,$dB stronger than the data SNR (i.e., $P_{\mathrm{p}}=10\,P_{\mathrm{d}}$) by spreading the pilots over many sub-carriers, which is typically possible in practice. Specifically, we set the $\mathrm{SNR}_{\mathrm{p}} = -10\,$dB and $\mathrm{SNR}_{\mathrm{d}} = -20\,$dB.

We set $(\varphi_1,\phi_1) = (\pi/2,0)$ as the initial angle pair for generating the codebook $\mathcal{B}$ using Algorithm~\ref{alg:Selection}. The users are distributed around the RIS such that $\varphi \sim \mathcal{U}[-\pi/3,\pi/3]$, and $\phi \sim \mathcal{U}[-\pi/3,\pi/3]$. These parameters are used to generate the channel according to \eqref{eq:A-set}, where $\boldsymbol{\psi} = (\varphi, \phi, r)$ and $r$ will be specified later.  

The results are generated using Monte Carlo simulations containing 2000 independent channel and noise realizations.
\begin{figure}
    \centering
    \includegraphics[width = 0.9\linewidth]{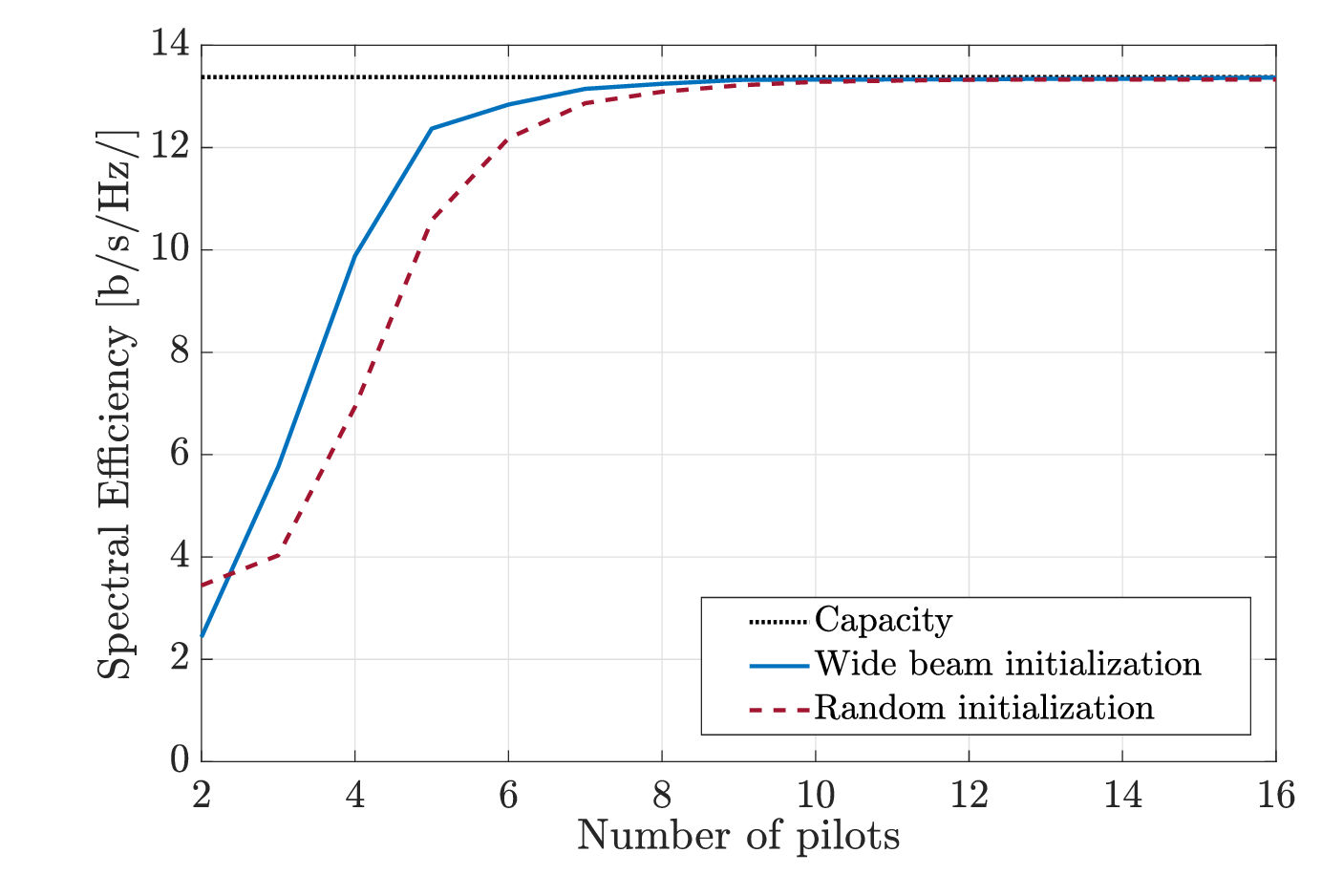}
    \caption{The average SE versus pilot length. The SE reaches the channel capacity with fewer pilots if we initialize Algorithm \ref{Alg:ML_Estimation} using wide beams instead of random narrow beams.}
    \label{fig:widecomp}
\end{figure}
\begin{figure}
    \centering
    \includegraphics[width = 0.9\linewidth]{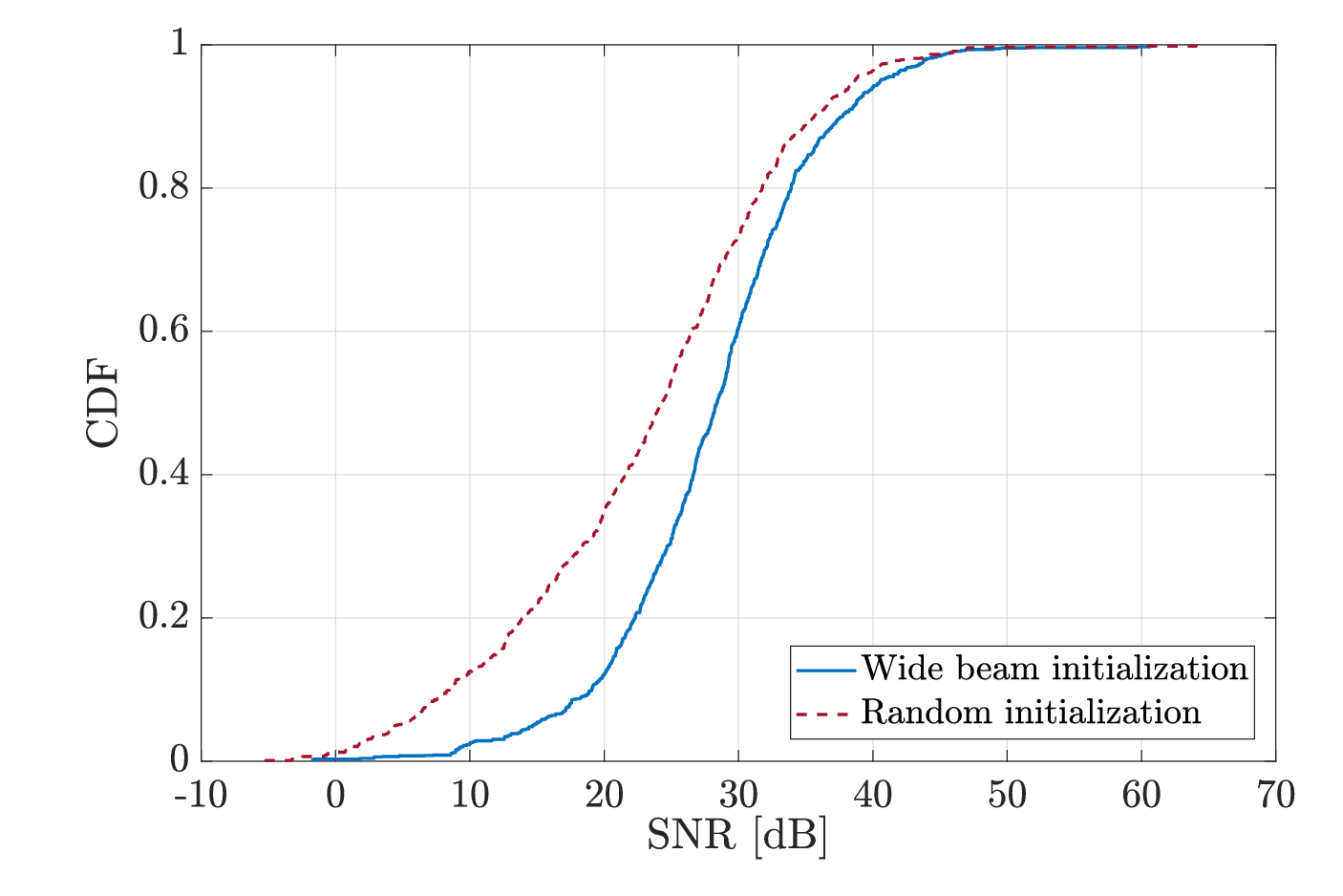}
    \caption{The CDF of the SNR at the BS during the first two pilot transmissions for random user locations. The wide beam initialization result in higher SNRs than random pilot initialization.}
    \label{fig:cdf}
\end{figure}
 \subsection{Convergence}
 \label{sec:convergence}
 We first examine the convergence of our proposed channel estimation algorithm in terms of the number of pilots required for the SE to reach capacity (achieved with perfect CSI). 

Fig.~\ref{fig:widecomp} shows the average SE versus the number of pilot transmissions.
Here, two initialization schemes are compared: the solid line uses the wide beam configurations from Section~\ref{subsec:widebeam-design} and the dashed line corresponds to the random initialization, where we pick two random columns of the codebook $\Theta$ to start the algorithm with the corresponding narrow beams. In both cases, we reach the capacity when several pilots are transmitted, but we observe faster convergence when the initialization is based on wide beams. The reason is that the wide beams provide useful information at the receiver irrespective of the user location, while the random narrow beams mostly do that when the main lobe or largest sidelobes happen to point towards the user location.
To further validate the efficiency of the wide beam initialization, Fig.~\ref{fig:cdf} depicts the cumulative distribution function (CDF) of the SNR at the BS for the first two pilot transmissions. The curve with the wide beam initialization starts at a larger SNR value and continues to outperform the random initialization scheme at all percentiles. There are specific user locations where it can be preferable, but statistically, the wide beam initialization scheme is superior. Hence, we initialize Algorithm \ref{Alg:ML_Estimation} using this scheme in subsequent simulations. 

\begin{figure}
    \centering
    \includegraphics[width = 0.9\linewidth]{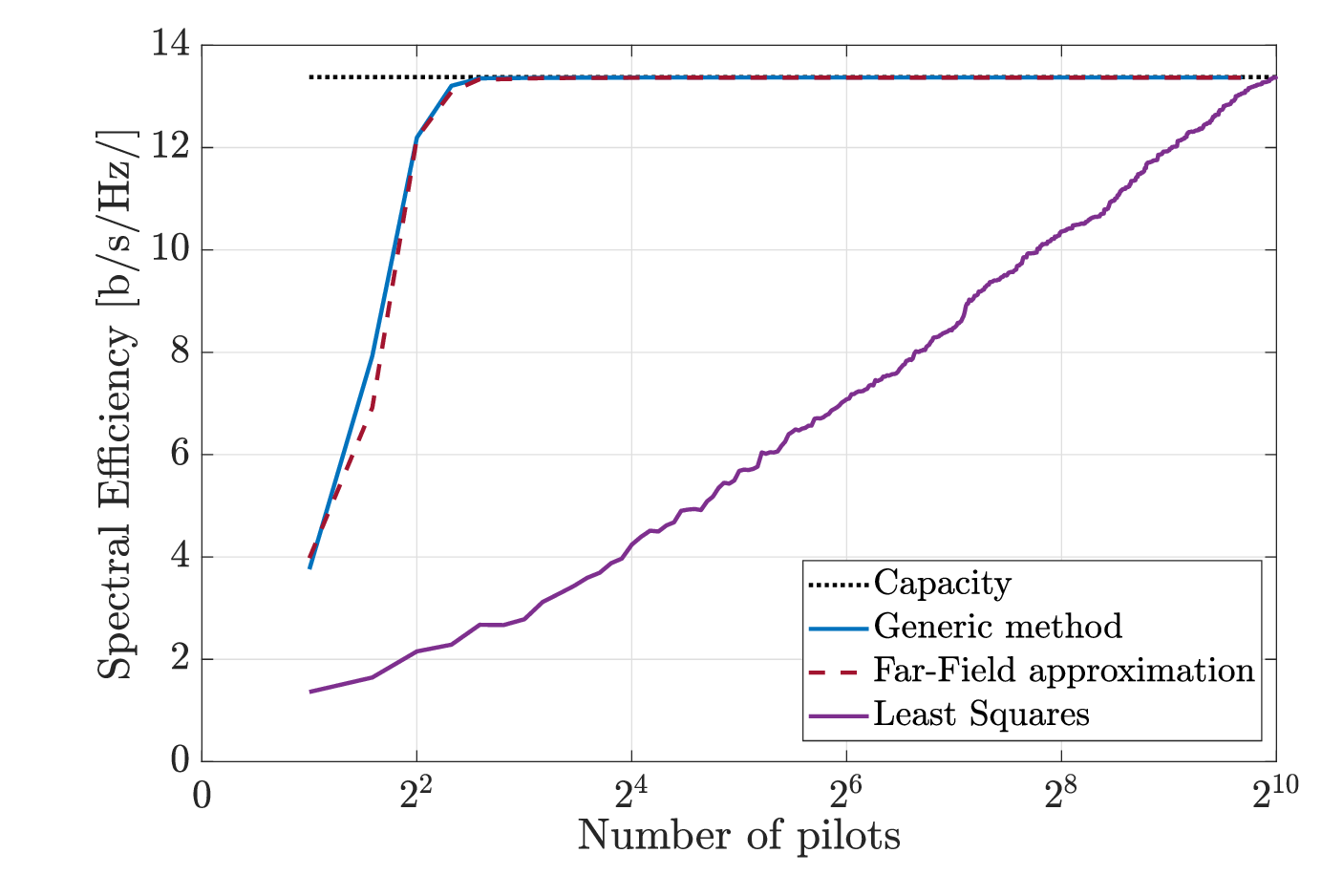}
    \caption{The average SE versus the pilot length when the user is at a random location in the far-field of the RIS.} \vspace{-2mm}
    \label{fig:FarField}
\end{figure}

\begin{figure}
    \centering
    \includegraphics[width = 0.9\linewidth]{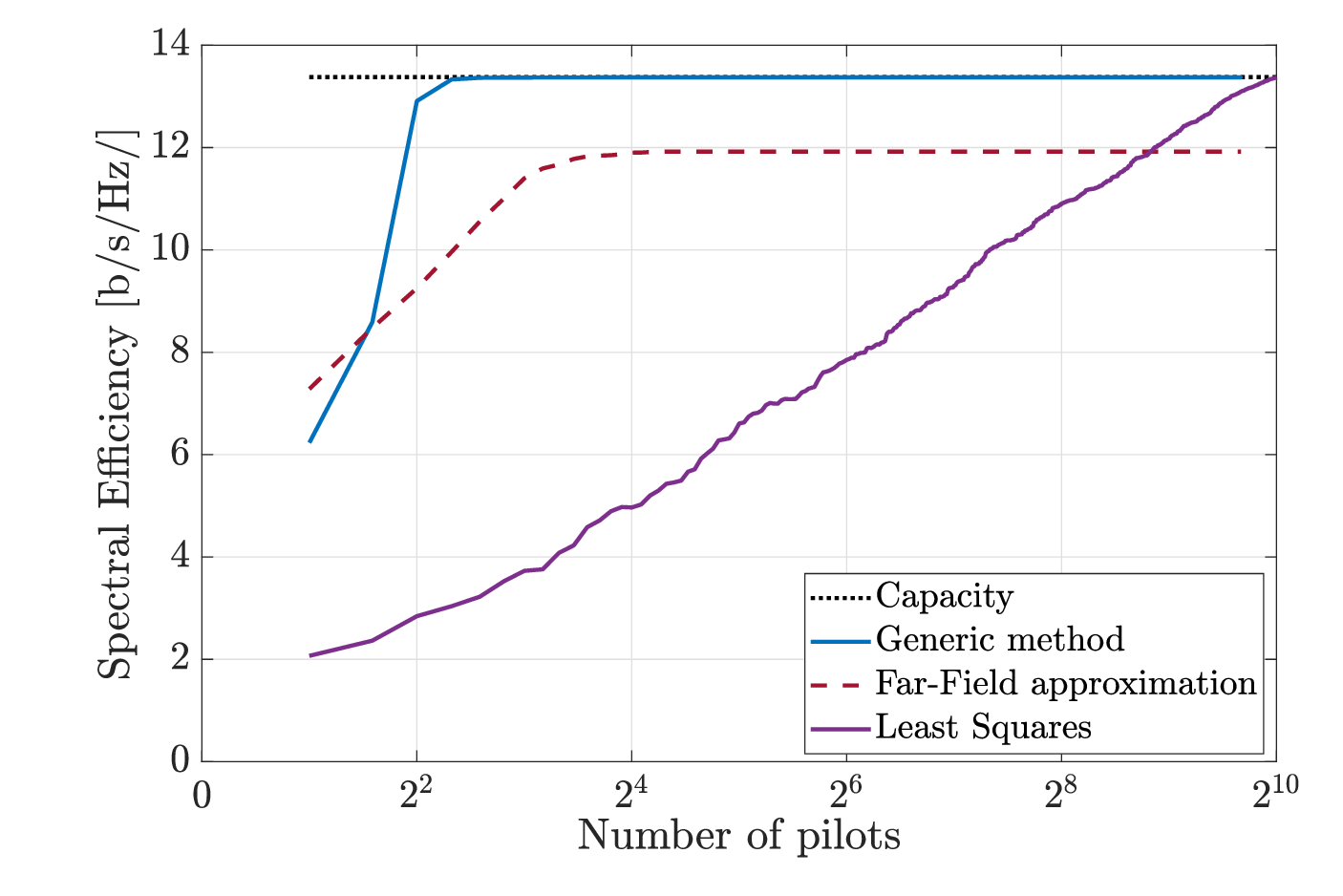}
    \caption{The average SE versus the pilot length when the user is at a random location in the near-field of the RIS.} \vspace{-2mm}
    \label{fig:NearField}
\end{figure}

Figs.~\ref{fig:FarField} and \ref{fig:NearField} evaluate the number of required pilots when the user is located in the far-field and near-field of the RIS, respectively. For the former case, the user distance is distributed as $r \sim \mathcal{U} [d_{\mathrm{FA}}/10,10\,d_{\mathrm{FA}}]$, and for the latter scenario $r \sim \mathcal{U}[d_{\mathrm{B}},d_{\mathrm{FA}}/10]$, where $d_{\mathrm{B}}$ is the Bj{\"o}rnson distance characterized as $d_{\mathrm{B}} = 2W$.\footnote{The Bj\"ornson distance is the distance beyond which the amplitude variations over the array is negligible \cite{ramezani2022bookchapter}, as is also assumed in this paper.} The solid blue curves correspond to the proposed algorithm when the exact generic array response model in \eqref{eq:array_response_general} is utilized in the MLE algorithm, while the red dashed curves show the case where the far-field approximation of the array response vector in \eqref{eq:array_response_farfield} is utilized. We can observe from Fig.~\ref{fig:FarField} that both methods converge to the capacity using only $8$ pilot transmissions, even if there are $1025$ complex channel coefficients to estimate; that is, only $0.78\%$ of the pilot resources required by non-parametric estimators. We stress that a logarithmic scale is used on the horizontal axis.
We notice that the far-field approximation can be used when estimating the channel of a far-field user, but it does not hurt to use the exact generic model. When the user is in the near-field region of the RIS, Fig.~\ref{fig:NearField} demonstrates that our algorithm with the exact generic array response model effectively estimates the channel and converges to the capacity using only $7$ pilots.
However, if a mismatched far-field channel estimator is utilized, 
we experience a loss of almost $1.5\,$bps/Hz in SE. The mismatch results in a performance ceiling when the number of pilots increases. Hence, it is essential to take near-field effects into account, and we should do it even if most users are in the far-field because the exact model gives correct results in both the near- and far-field regions. 

In the same figures, we compare our proposed method with a classical non-parametric estimator: least squares (LS). The LS utilizes the columns of the $1025 \times 1025$ DFT matrix as RIS configurations during the pilot transmission. We observe that our parametric method significantly outperforms LS in both far-field and near-field scenarios, except when the $1025$ pilots are used. This demonstrates that conventional non-parametric methods require $N+1$ pilots to obtain a useful estimate.

\subsection{User Tracking}
Having verified the effectiveness and efficiency of the proposed channel estimation scheme in estimating a static user-RIS channel, we now proceed to evaluate its performance in estimating a time-varying channel. We consider an indoor scenario with non-stationary mobility; that is, the user-RIS channel is not characterized by time-correlated small-scale fading from a stationary distribution as in \cite{Papazafeiropoulos2022a,Zhang2022a,Xu2022,Chen2022a} but random large-scale effects (i.e., angles and distances).

\begin{figure}
    \centering
    \includegraphics[width = 0.9\linewidth]{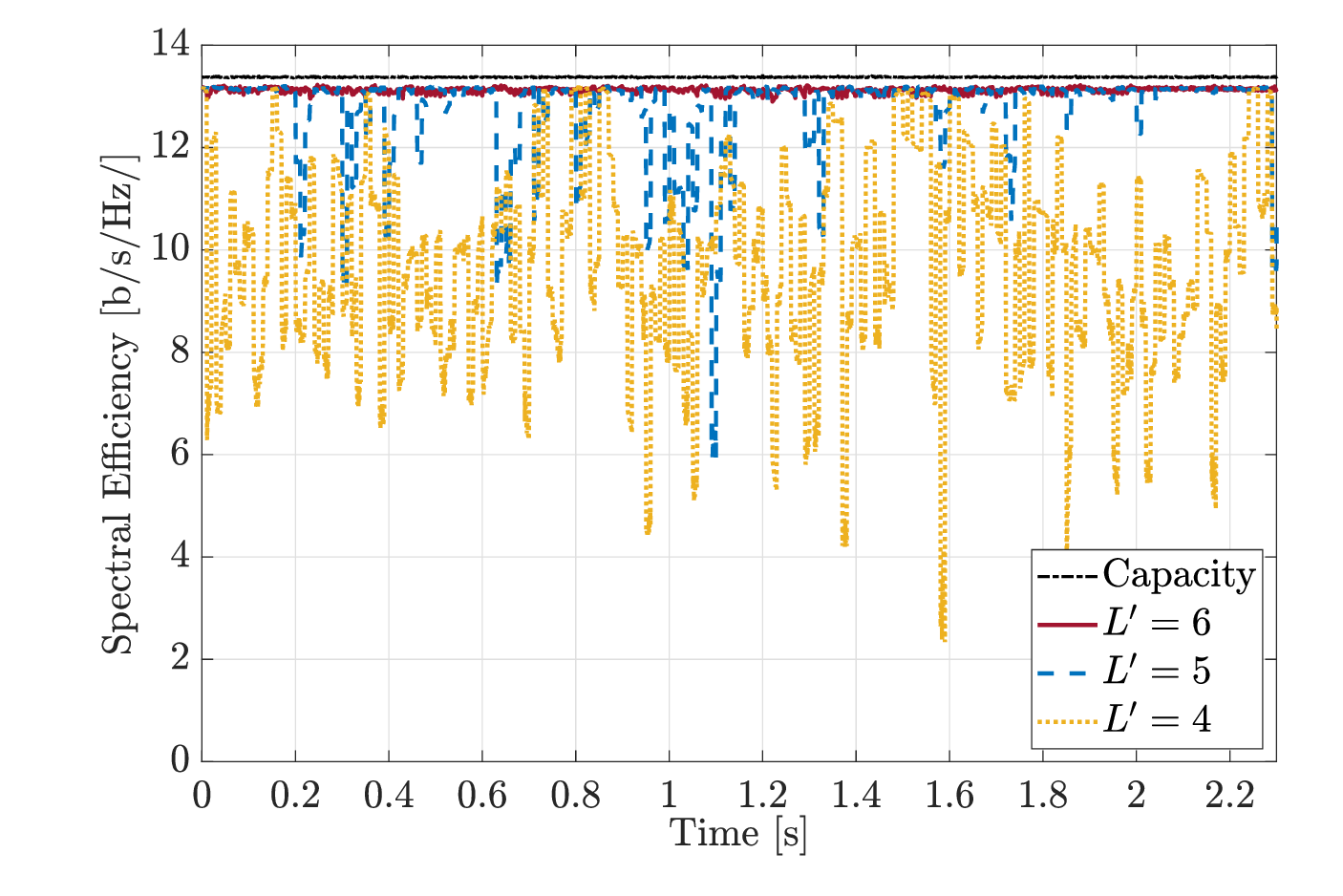}
    \caption{A user walks inside a room experiencing both far-field and near-field channel conditions. The initial channel estimation is performed using $L = 10$. For the subsequent channel re-estimations, the MLE scheme uses $L'$ pilot symbols.}
    \label{fig:RandomWalk}
\end{figure}

We consider an indoor RIS-assisted communication scenario  where a user walks within a $4\,$m $\times4\,$m room with the average speed of $1\,$m/s and its minimum distance to the RIS is $d_{\mathrm{B}} = 0.48\,$m. In this setup, about $10\,\%$ of the room falls in the radiative near-field region of the RIS. The user is assumed to be at the same height as the center of the RIS and the movement direction is uniformly distributed over $[0,2\pi]$. Due to the user movement, the channel between the user and the RIS and the direct channel experience variations over time,  necessitating the re-estimation of the channel and re-configuration of the RIS phase shifts. Fig.~\ref{fig:RandomWalk} depicts the achievable SE over time, where the channels and SE are computed once per millisecond.
The channel estimation is performed every $10\,$ms using the following approach: We start the MLE channel estimation with $L=10$ pilots using the wide configurations in \eqref{eq:initial_RIS}. For the remaining channel estimation phases, we set the number of pilots to $L^\prime$ and use a smart initialization strategy where the two initial RIS configurations $\{\boldsymbol{\theta}_1,\boldsymbol{\theta}_2\}$ are selected based on the last successful RIS configuration for data transmission. 
Specifically, if $\bar{\boldsymbol{\theta}}$ is the final RIS configuration after fully estimating the channel, we select $\{\boldsymbol{\theta}_1,\boldsymbol{\theta}_2\}$ as
\begin{align}
\label{eq:smartInit}
    \boldsymbol{\theta}_1 = \argmax{\boldsymbol{\theta} \in \Theta}~ | \bar{\boldsymbol{\theta}}^{\Htran} \boldsymbol{\theta}|,\\
    \boldsymbol{\theta}_2 = \argmax{\boldsymbol{\theta} \in \Theta\setminus{\{\boldsymbol{\theta}_1}\}}~ | \bar{\boldsymbol{\theta}}^{\Htran} \boldsymbol{\theta}|,
\end{align}
where $\Theta$ is the codebook generated by Algorithm~\ref{alg:Selection}. We can see in Fig.~\ref{fig:RandomWalk} that there are significant drops in the value of SE when only $4$ pilots are used for channel re-estimation. The SE improves when $L^\prime = 5$, but the algorithm still cannot fully track the channel variations because of the random motion. However, with $L^\prime = 6$ pilots, the proposed MLE framework together with the smart initialization strategy can effectively track the user and keep the SE close to its maximum value. Hence, if we have prior information about the user's location, we can use it instead of wide beams to further speed up the convergence of the proposed MLE algorithm.

\vspace{-2mm}

\subsection{Accuracy under Rician Fading }

We have so far demonstrated the effectiveness of the proposed parametric MLE method in estimating a pure LOS channel between the user and the RIS.
We now extend the evaluation of our presented framework to the more practical Rician fading scenario where in addition to the dominant LOS link, there also exist NLOS channel components. Specifically, the user-RIS channel is modeled as:
\begin{figure}
    \centering
    \includegraphics[width = 0.9\linewidth]{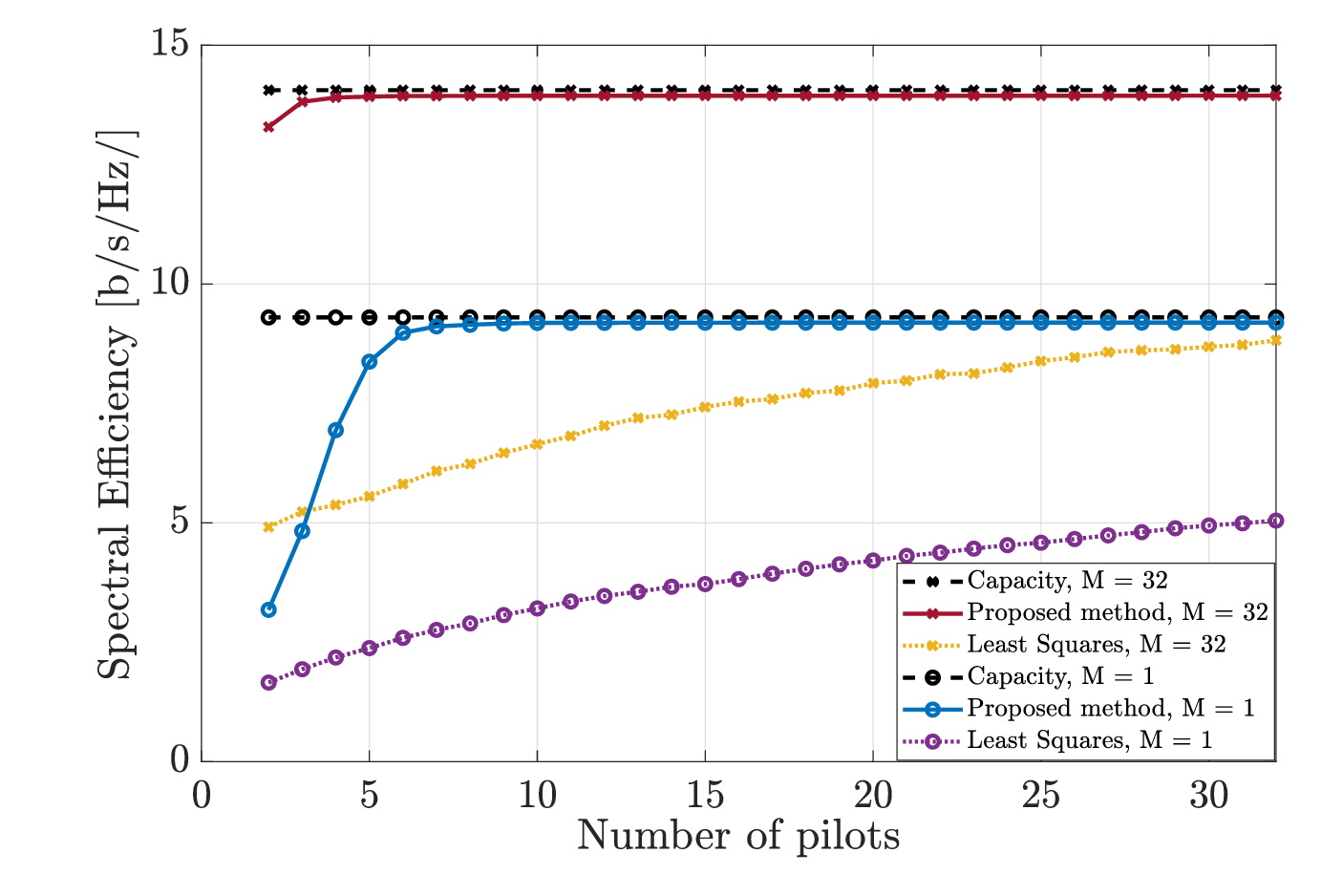}
    \caption{The average SE versus the pilot length when the channels follow Rician fading with the K-factor of $8\,$dB.}
    \label{fig:SE-SIMO-VS_SISO}
\end{figure}

\begin{align}
\label{eq:g-rician}
    \vect{g} &=  \sqrt{\frac{K}{K+1}}\vect{g}_{\mathrm{LOS}} + \sqrt{\frac{1}{K+1}}\vect{g}_{\mathrm{NLOS}},
\end{align}
where $K$ denotes the Rician factor defined as the ratio between the power of the LOS path and combined NLOS paths. The LOS term is generated according to~\eqref{eq:A-set}, while the NLOS term follows correlated Rayleigh fading~\cite{EmilMIMObook}. Other channels are modeled similarly and the Rician factors are set to $8\,$dB. We consider two setups at the BS, one with single antenna and the other with $M = 32$ antennas in the form of a uniform linear array (ULA) with half-wavelength inter-antenna spacing. The RIS has UPA structure with $16 \times 16$ elements and inter-element spacing of half a wavelength. In addition, we set $\mathrm{SNR_p} = -10\,$dB and $\mathrm{SNR_d} = -20\,$dB. 

Fig.~\ref{fig:SE-SIMO-VS_SISO} illustrates the SE as a function of pilot length in the two mentioned scenarios. Similarly to the observations of previous simulations, the convergence occurs
within a few pilot transmissions, endorsing the efficiency of the proposed MLE framework once again. Importantly, when there are multiple antennas at the BS, the convergence happens within three pilot transmissions which is fewer than the number of required pilots in case of a single-antenna BS. 
This stems from the fact that using multiple antennas allows the BS to acquire multiple observations of the channel from one pilot transmission, as a result of which, less number of pilots are needed for obtaining an accurate estimate of the channel compared to the single-antenna BS scenario.
Moreover, the proposed method demonstrates near-capacity performance under  Rician fading channel models. This is of particular importance since it shows that the proposed MLE method effectively works under practical mmWave channels with a relatively large K-factor \cite{2013mmWaveSayeed}, although we have used a parametric LOS  model for $\vect{g}$ in developing the presented method. 
As expected, the LS estimator requires many more pilots to obtain an accurate channel estimate.

\subsection{Normalized Mean Squared Error}
\begin{figure}
    \centering
    \includegraphics[width = 0.9\linewidth]{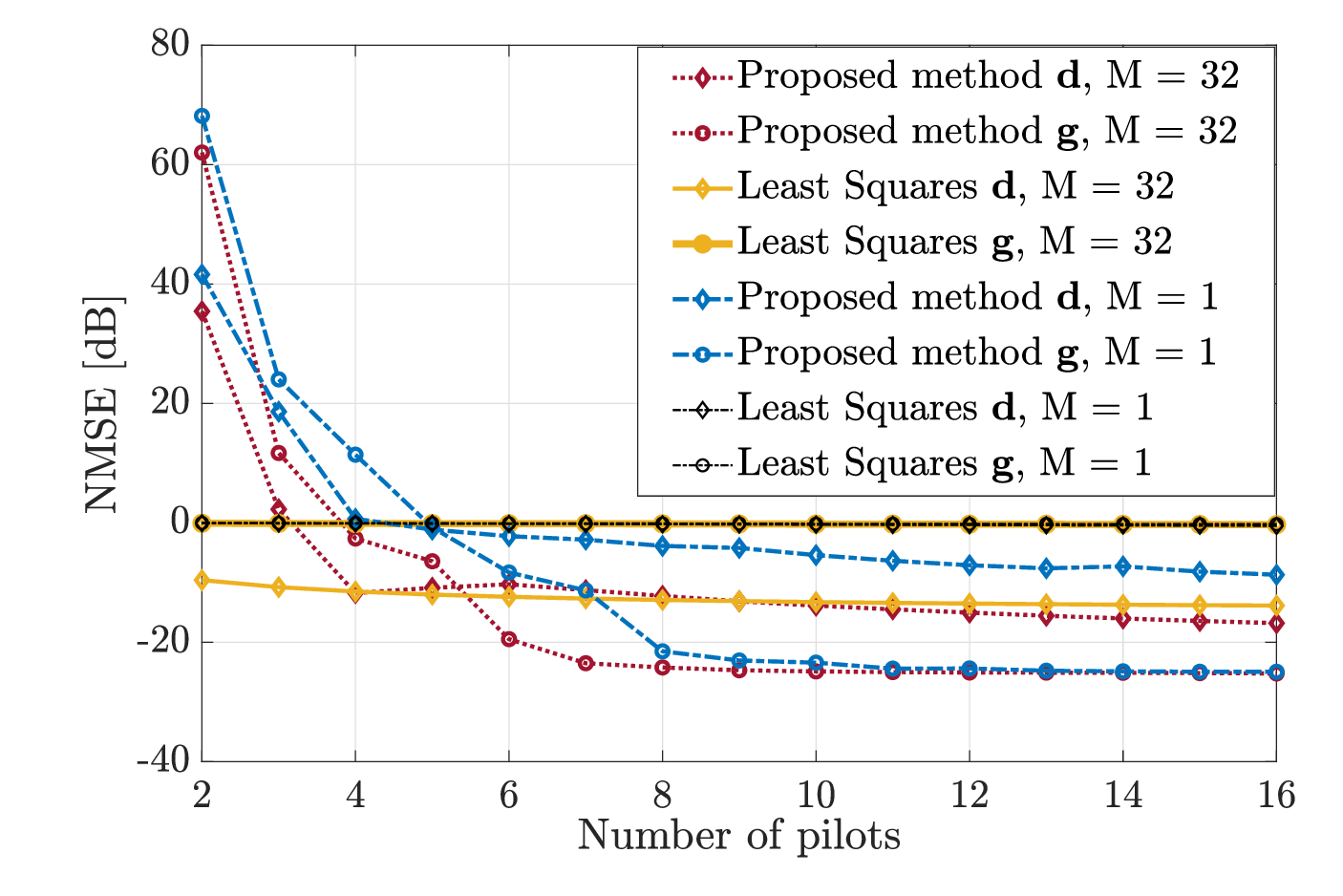}
    \caption{NMSE versus pilot length for the estimated channels.}
    \label{fig:NMSE-SIMO-VS_SISO}
\end{figure}

 Fig.~\ref{fig:NMSE-SIMO-VS_SISO} depicts the normalized mean squared error (NMSE) for the estimated channels. Again, both single-antenna and multi-antenna BS scenarios are considered where in the latter case we have $M = 32$. The NMSE for the estimated direct channel ($\vect{d}$) and the estimated channel between the user and the RIS ($\vect{g}$) can be calculated as follows:
\begin{equation}
    \mathrm{NMSE}_{\vect{d}} = \mathbb{E}\left\{ \frac{\| \hat{\vect{d}} - \vect{d}\|^2}{\| \vect{d}\|^2}\right\},
\end{equation}
\begin{equation}
    \mathrm{NMSE}_{\vect{g}} = \mathbb{E}\left\{ \frac{\| \hat{\vect{g}} - \vect{g}\|^2}{\| \vect{g}\|^2}\right\}.
\end{equation}
Here, the NMSE values are obtained through Monte Carlo simulations.
We can see that the NMSE for our proposed method generally decreases with the number of pilots for both $\vect{d}$ and $\vect{g}$ and in both scenarios which is due to the fact that the channel estimation accuracy improves with more pilot transmissions. Moreover, the accuracy of estimation is superior in scenarios involving a multi-antenna BS compared to those with a single-antenna BS. This difference arises from the fact that transmitting a single pilot symbol in the case of a multi-antenna BS is analogous to transmitting multiple pilots in the single-antenna BS scenario. In this figure, we additionally display the NMSE results for the LS estimator. The corresponding curves for the user-RIS and direct channels in the single-antenna case and the user-RIS channel in the multi-antenna case are almost flat for the number of pilots shown on the figure because the LS estimator requires more pilots to achieve an accurate estimate. 

\subsection{Comparison with Hierarchical Beam Training}
To further corroborate the effectiveness of the proposed adaptive RIS configuration strategy, we compare it with the well-known hierarchical beam training. Fundamentally, both methods follow similar policies which is to iteratively determine the optimal configuration within a codebook during pilot transmission. In our method, the selection of the phase shifts for next pilot transmission is based on the estimated channel using previously received pilots. On the other hand, the hierarchical search scheme initially configures the RIS such that the surface generates two wide beams. The received signal powers are compared, and the configuration which has led to the beam with a higher SNR is selected. For the transmission of the next two pilots, RIS configurations are designed to produce two narrower beams within the previously selected wide beam. Again, the configuration resulting in a higher SNR is picked. This process is repeated until the narrowest beam in the codebook is identified and the corresponding RIS configuration is selected as the optimal RIS configuration during the data transmission phase. 

\begin{figure}
    \centering
    \includegraphics[width = 0.9\linewidth]{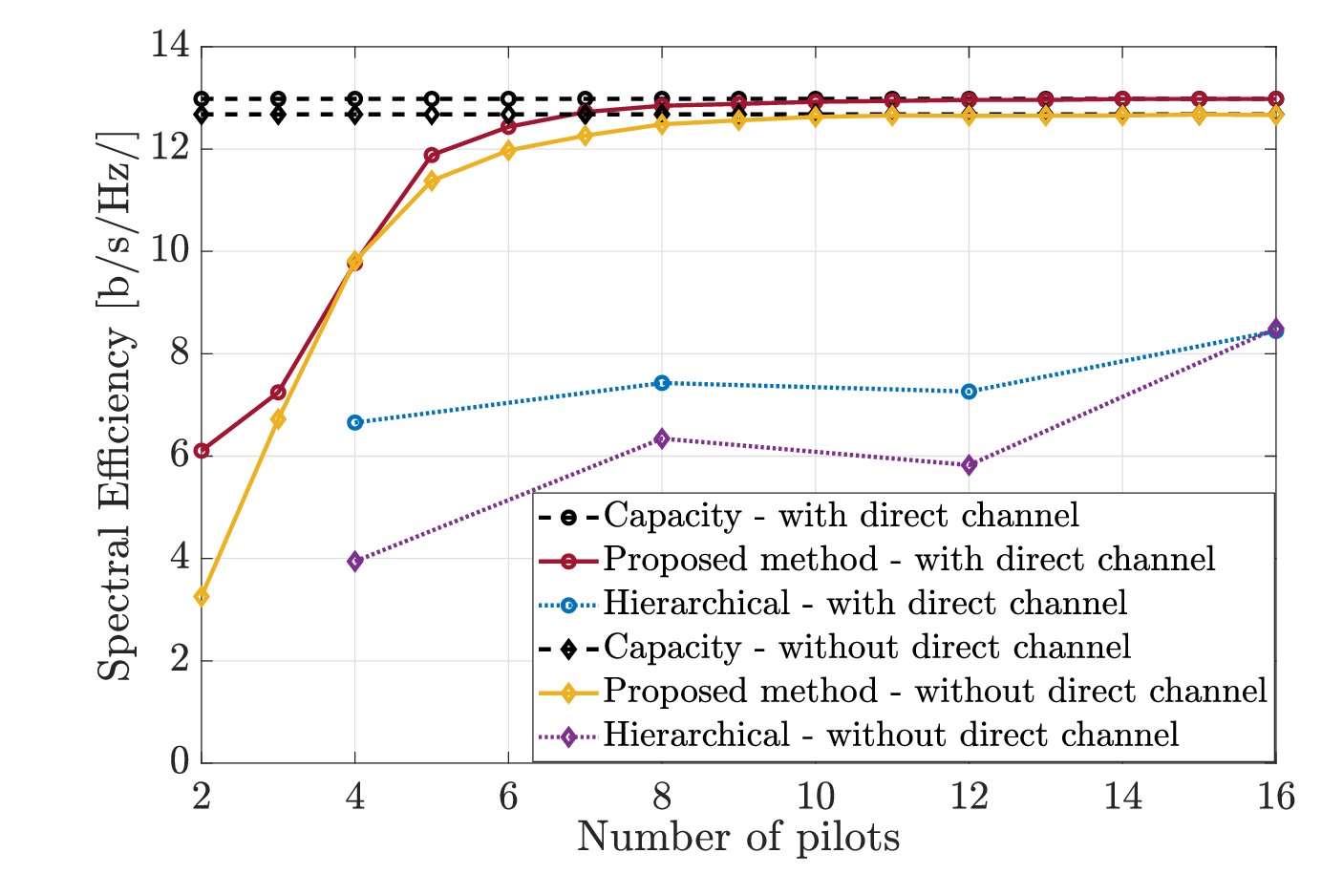}
    \caption{The average SE versus the pilot length for the proposed method and the hierarchical beam training method.}
    \label{fig:SE-Hierarchical}
\end{figure}

We use the hierarchical codebook design in ~\cite{Hierar2016} 
 as the benchmark. 
 The RIS has a UPA structure with $16 \times 16$ elements. The hierarchical beam search  presented in \cite{Hierar2016} has thus been modified for the UPA structure. Since the array response vector of a UPA can be represented as the Kronecker product of those of two ULAs, the binary-tree structure of the search turns into a quaternary-tree one, where in each iteration of the algorithm, four RIS configurations are examined and the one resulting in the highest SNR is taken as the RIS configuration for the next pilot transmission.
 The BS has one single antenna and we have $\mathrm{SNR}_{\mathrm{p}} = 0\,$dB and $\mathrm{SNR}_{\mathrm{d}} = -10\,$dB. We consider two scenarios where in the first one, the direct channel between the BS and the user is obstructed and in the second setup, the existing direct link is $64$ times stronger than the cascaded path via a single RIS element. The number of pilot symbols is limited to $4\cdot\log_2(16) = 16$, representing the overhead required for the hierarchical scheme to reach the narrowest beam in the codebook. 
Fig.~\ref{fig:SE-Hierarchical} depicts the performance of the proposed method and the hierarchical beam training method in converging to the capacity. It can be clearly seen that the hierarchical beam training scheme fails to reach the capacity because it is not able to accurately identify the location of the user. On the other hand, the proposed scheme quickly converges to the capacity since the employed adaptive RIS configuration strategy allows for effectively estimating the channel parameters with a few pilot transmissions. 

\section{Conclusions and Future Work}
\label{sec:conclusion}
Parametric channel estimation methods can use a known channel structure to shorten the pilot length. In an ideal noise-free scenario, the minimum pilot length equals the number of parameters to be estimated. However, parametric methods can be unreliable at practical SNRs since the pilots only excite a tiny subset of the vector space where the channel resides. In this paper, we proposed a novel parametric MLE framework for LOS channel estimation in RIS-aided systems. We circumvent the aforementioned issue by adaptively adjusting which RIS configurations are used during pilot transmission to gradually refine the estimation accuracy. We devised a minimal-sized codebook of orthogonal RIS configurations to choose from during pilot transmission. To further reduce the average number of required pilots needed to reach capacity, we designed two wide beam RIS configurations to initialize the MLE algorithm with. This ensures that useful information about the user’s channel parameters can be extracted even if the SNR is generally low, thereby reducing the total number of pilots accordingly. The effectiveness of the proposed design was verified via numerical simulations. In both near- and far-field scenarios, the pilot length could be reduced from $1025$ (required by non-parametric methods) to $8$. The number can be further reduced to $6$ in user mobility scenarios, where there is prior information regarding the user's previous location.

For future work, a promising direction is to extend the proposed MLE framework to wideband scenarios where the channels are frequency-selective. Furthermore, taking into account the phase-dependent amplitude variations of RIS elements during adaptive RIS configuration is another important topic to explore. 
\vspace{-3mm}
\bibliographystyle{IEEEtran}
\bibliography{IEEEabrv,refs}

\end{document}